\documentclass[%
reprint,
 amsmath,amssymb,
 aps,
 prb,
floatfix,
]{revtex4-2}

\usepackage{graphicx}
\usepackage{dcolumn}
\usepackage{bm}
\usepackage{braket}

\begin{document}

\preprint{APS/123-QED}

\title{Realizing attractive interacting topological surface fermions: A resonating TI- thin film hybrid platform}

\author{Saran Vijayan and Fei Zhou}
\affiliation{Department of Physics and Astronomy, University of British Columbia, 6224 Agricultural Road, Vancouver, BC, V6T 1Z1, Canada}
\begin{abstract}
In this article, we propose a practical way to realize topological surface Dirac fermions
with tunable attractive interaction between them. The approach involves coating the surface of a topological insulator with a thin
film metal and utilizing the strong-electron phonon coupling in the
metal to induce interaction between the surface fermions. We found that for a given TI and thin film, the attractive interaction between the surface fermions can be maximally enhanced when the Dirac point of the TI surface resonates with one of the quasi-2D quantum-well bands of the thin film. This effect can be considered to be an
example of ’quantum-well resonance’. We also demonstrate that the superconductivity of the resonating surface fermions can be further enhanced by choosing a strongly interacting thin film metal or by tuning the spin-orbit coupling of the TI. This TI-thin film hybrid configuration holds promise for applications in Majorana-based quantum computations and for the study of quantum critical physics of strongly attractively interacting surface topological matter with emergent supersymmetry. 
\end{abstract}

\maketitle

\section{\label{sec:level1}Introduction}
Topological Insulators(TI)\cite{fukane(2005),bernevigzhang(2006),bernevighugheszhang(2006),fukanemele(2007),moorebalents(2007),hasankanereview(2010), Qizhangreview(2011),bernevigbook(2013)} belong to the class of symmetry-protected topological phases, where the gapless boundary states are protected by Time-Reversal Symmetry(TRS). One interesting feature of these surface states is that their low-energy excitations can resemble a single flavor of 2-component massless Dirac fermions($N_f = 1/2$). This is unique because it is impossible to realize an odd number of flavors of 2-component Dirac fermions in a bulk lattice because of the fermion doubling problem intrinsic to lattice models\cite{Nielson(1981),Nielson(1981)(2)}. Therefore topological surface provides an interesting platform to study various interactions involving a single flavor of 2-component Dirac fermions, provided the interactions do not break the time-reversal symmetry.

Of particular interest is when there is an effective attractive interaction between the surface fermions. For a finite chemical potential (i.e. when the Fermi level is above or below the Dirac point), the U(1) symmetry breaking leading to the superconducting phase can happen for arbitrarily weak attractive interaction due to the Cooper instability at the surface. On the other hand at the zero chemical potential (Fermi level aligned with the Dirac point), the interaction strength must be greater than a critical value for the phase transition to occur. In both these cases, the resulting superconducting phase can be of non-trivial topological character\cite{Jackiw(1981),fukane(2008)}. Specifically, it implies that the vortex core of the superconductor can host Majorana zero modes\cite{Read(2000),Alicea(2012),beenakker(2013),sato(2016),sato(2017)}. They are considered to be an ideal candidate for fault-tolerant quantum computing because of their non-Abelian statistics.

Another interesting feature of the attractively interacting surface fermions is that surface dynamics have an emergent Lorentz symmetry when the chemical potential is zero. It has been demonstrated that the effective field theory of the surface states further exhibits emergent supersymmetry (SUSY) when the coupling constant of the attractive interactions is tuned to be quantum critical \cite{SSLee(2007),Grover(2014),Ponte(2014),yao(2017),yao(2017)(2),yao(2018)}. Supersymmetry is the symmetry between bosons and fermions and had been speculated to exist as a fundamental symmetry in elementary particle physics. Emergent supersymmetry in lattice models is difficult to realize, at least in $d>1$ spatial dimensions, because fermions typically have more degrees of freedom than bosons in lattices, a consequence of the fermion doubling problem. But at a quantum critical point of topological surfaces, an emergent SUSY  exists between the charge {\em 2e} bosons that naturally emerge as quasi-particles and the 2-component Dirac fermions in the semi-metallic phase, both of which can be strongly self-interacting and mutually interacting. Therefore, the topological surface provides an ideal platform to study the dynamics of supersymmetric quantum matter.

However, realizing a topological surface with net attractive interactions between them is not straightforward and can be challenging. One reason is the unscreened nature of repulsive Coulomb interactions in an insulator. In addition, many topological insulator materials do not have strong electron-phonon interactions. In this article, we propose coating the 3D TI surface with a metallic thin film as a practical way to realize a ground state of interacting surface fermions with net attractive interaction between them. A thin film is characterized by the quasi-2-dimensional quantum-well bands due to the quantum confinement of the electronic states in the third dimension. Due to the screened nature of Coulomb repulsion, the phonon-mediated attractive interaction can be the dominant form of interaction between electrons at zero temperature. On depositing the thin film to the TI surface, the 2D surface Dirac fermions and the quasi-2D quantum-well fermions start hybridizing. These hybrid fermions are a quantum superposition of the quantum-well states and the TI surface states. Hence the hybrid fermions not only acquire a helical spin-texture from the surface side but will also experience a net attractive interaction due to coupling with the phonons in the thin film. In a way, hybridization causes the surface Dirac cone to be exported to the thin film which results in the helical Dirac fermions experiencing a phonon-mediated attractive interaction between them. Alternatively, one can show that the hybridization leads to variable or tunable attractive interactions among topological surface Dirac fermions.

We have observed that this attractive interaction between the helical fermions is maximally enhanced when the Dirac point of the TI surface resonates with one of the quantum-well states of the thin film. While at resonance, there is no clear distinction between the TI surface and the thin film states as the electronic states are strongly hybridized, we do show that in the wide range of parameter space, the low energy physics effectively becomes that of strongly interacting surface Dirac fermions.  We study the superconductivity of these resonating hybrid states at different thickness regimes. Consider the ultra-thin limit of the film, when only a single quantum-well (QW) band crosses the Fermi level (we shall call this the $N = 1$ limit, where $N$ is the number of QW bands crossing the Fermi level). Then we effectively have a four-band model of the interacting helical hybrid states. Following the bulk-boundary relations(BBR) of interactions obtained before\cite{saran(2022)}, we find that effective phonon-mediated interaction scales as $1/D$, $D$ being the thickness of the film and hence the interaction strength is at its strongest in the $N=1$ ultrathin limit. We show that for a wide range of chemical potentials, it is possible to construct an effective field theory of attractively interacting 2-component Dirac fermions. We then studied possible ways of enhancing the superconducting gap by tuning the bulk coupling constant of the thin film and the Dirac velocity of the surface fermions.

When the thin film thickness is increased further, in addition to the Fermi surfaces formed by the resonating hybrid bands, there also exists Fermi surfaces formed by the QW bands that were off-resonance. Therefore the superconducting gap in this limit is formed not just due to attractive interaction between the surface fermions but also because of the scattering of the singlet pair of electrons from these background off-resonance QW Fermi surfaces. In the very thick limit (large-$N$ limit), we explicitly show that the superconductivity on these resonating hybrid bands is dominated by the scattering of the singlet pair of electrons from the off-resonance Fermi surfaces. However, the interaction between the surface fermions can further enhance the surface superconductivity.  And when the interactions are sufficiently strong, the enhancement can be very substantial.

It should be noted here that if one's prime focus is mainly to realize a topological superconducting phase, then it is not necessary to have attractive interaction between the surface fermions\cite{fukane(2008)}. Superconductivity can be induced on the surface by the proximity effect, implemented by depositing a bulk s-wave superconductor on the TI surface. The interface between the TI and the s-wave superconductor had been shown to be in the topological superconducting phase even though the surface electrons are non-interacting. As mentioned before, the main objective of our work is to realize a platform of strongly interacting surface fermions. The attractive interactions between surface fermions can lead to emergent SUSY at its QCP, a phenomenon that can potentially have a lot of impacts on the fundamental understanding of the building blocks of nature.

However, the TI-thin film hybrid has richer physics over the conventional proximity structures even if our objective is only to realize a topological superconducting phase. Due to the strong single-particle hybridization, there is a finite probability of finding the surface fermions on the thin film side, sometimes called the 'topological proximity effect' \cite{stanescu(2011),chiang(2013),Shoman2015}. Thus in this structure, the topological superconducting phase can proliferate across the interface, and can even be observed on the thin film side and not just at the interface, making it easy to detect in the experiments \cite{Trang2020}.

We like to note here that the effect of tunneling of the TI surface fermions on the superconductivity {\em in the thin film} has been extensively studied in refs.\cite{sudbo(2019),sedlmayr(2021),liu(2022)}. Ref.\cite{sudbo(2019)} studied the superconductivity in the monolayer thin film-TI hybrid as a function of tunneling strength and the chemical potential.  They found a suppression of the superconducting gap in the thin film when the Fermi momentum of the thin film and the TI surface matched and non-hybridized surface fermions were integrated out.
Ref.\cite{liu(2022)} found an enhancement in the superconducting order when the Fermi level crosses the bottom of the double-well hybrid bands. This result is encouraging in the context of thin film superconductivity. However, the Lifshitz transition leading to the enhancement results in two additional fermion surfaces and does not affect the topological aspect of superconducting pairing.

The main focus of this article on the other hand is to understand Dirac fermions in the topological surface and their interactions and pairing dynamics mediated by coated thin films.
When the superconductivity of non-interacting surface fermions (before the tunneling is turned on) is concerned, we find in this work that the superconductivity on the surface fermions can be induced and greatly enhanced if surface fermions are in resonance with electrons in thin films.
Although in the limit of resonance physically it is not possible to entirely isolate the surface fermions from the thin film electrons, the effective field-theory description in the most interesting limit is simply of the form of interacting Dirac fermions but with various substantially renormalized parameters. These renormalization effects especially the fermion-field renormalization are one of the main focuses of our studies and discussions below as they directly set the strength of interactions mediated by the thin films. These renormalization effects can either lead to surface superconductivity that otherwise won't exist because of the absence of direct pairing dynamics or further enhance the well-known Fu-Kane proximity effects of non-interacting surface fermions\cite{fukane(2008)}. The induced surface fermion interactions are also shown to follow explicitly the generic scaling law indicated in the general bulk-boundary interaction relation obtained in a previous article\cite{saran(2022)}.

The article is organized as follows: In the section \ref{sec:level2}, we discuss the single particle tunneling physics at the interface. We write down the single-particle Hamiltonian for the hybrid fermions on the helicity basis. In section \ref{sec:level3}, starting from the fundamental electron-phonon coupling Hamiltonian of the thin film, we derive a general short-ranged pairing Hamiltonian that explains the interactions of the hybrid fermions with one another and with the thin film electrons belonging to the off-resonance bands. Here we find that the hybrid fermions acquire an effective attractive interaction between them and the interaction strength is renormalized by a $Z$-factor. The $Z$ factor is essentially a measure of the probability amplitude of the hybrid fermions to be in the thin film side of the interface. In section \ref{sec:level4}, we study the evolution of this $Z$-factor as a function of the dimensionless detuning parameter $\tilde{\delta}$ and find that the attractive interaction between the surface fermions is enhanced at the quantum-well resonance ($\tilde{\delta} = 0)$. Section \ref{sec:level5} is dedicated to the mean-field approximation. Here we derive the superconducting gap equation under the assumption that the Debye frequency $\omega_D \ll \mu$, where $\mu$ is the chemical potential. Essentially, we assume that only electronic states near the Fermi level are interacting.  In section \ref{sec:level6}, we consider the limit when the surface states hybridize with the $N=1$ QW band of the thin film. We construct an effective theory followed by exploring various ways to enhance the superconducting gap in this limit. Section \ref{sec:level7} discusses the large-$N$ limit of the theory. Here we make connections to Fu-Kane's model in the perturbative limit of tunneling. In Section \ref{sec:level8}, we studied the evolution of the superconductivity on the resonating hybrid states as a function of thickness (parametrized by the band index $N$).  
\begin{figure*}
\includegraphics[width=5.94cm, height=4.52cm]{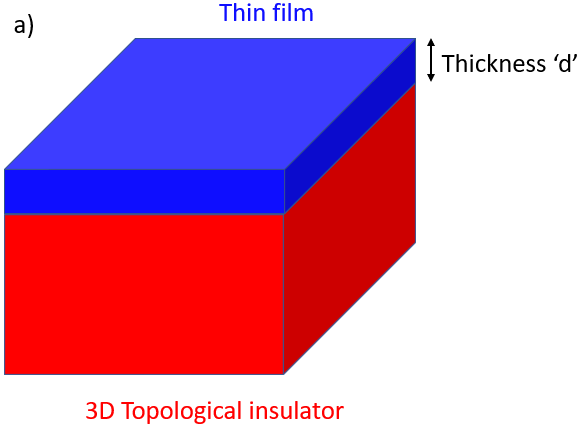}
\includegraphics[width=11.7cm, height=7.40cm]{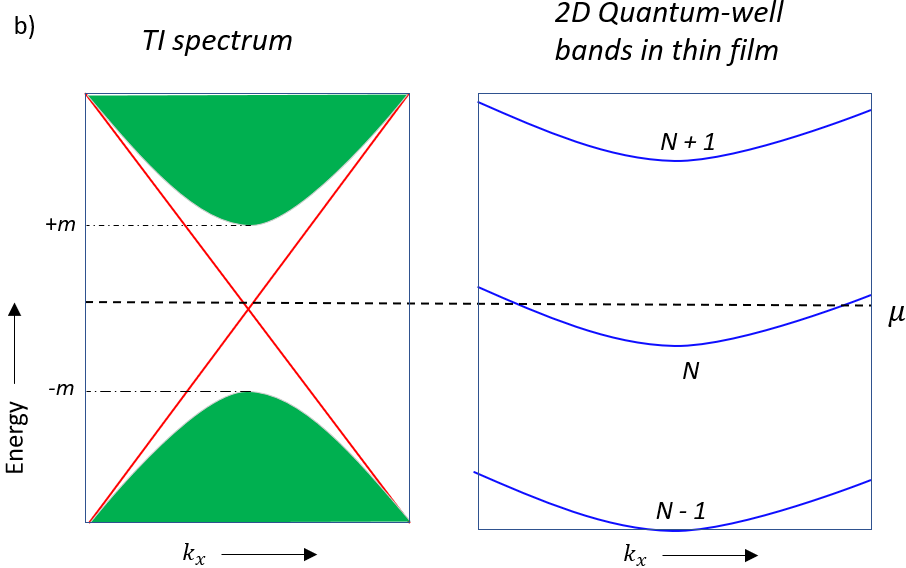}
\caption{\label{TI-thfmschematic} a)TI-thin film hybrid considered in this paper. b)Left: Schematic picture of the surface states of the topological insulator(Red lines). The gapped bulk bands are indicated by the green-shaded area. Here, $m$ stands for the bulk energy gap. Right: Schematic picture of the effective 2D quantum-well bands of the thin film considered in this article. The chemical potential $\mu$(dashed line) is set within the bulk energy gap of the TI. We only consider the case when the $N$th band that is energetically closest to the Dirac point of the TI surface is separated from the two adjacent QW bands by an energy separation much greater than the $m$, the bulk gap of the TI (See Eqn.\ref{condition}). In this limit, the tunneling effects on the $N\pm1$ QW bands from the surface Dirac cone are negligible and can be conveniently ignored.}
\end{figure*}
\section{\label{sec:level2}Non-interacting theory}
\subsubsection{Model Hamiltonian}
We start by defining a minimal theoretical model to understand the essential tunneling physics at the thin film-topological insulator interface.  Let the thin film - TI interface be at $z = 0$. The topological insulator(TI) occupies the bottom half-plane defined by $z < 0$. Consider a thin film of thickness $d$ deposited over the TI surface, so that it occupies the space $0 < z < d$. 

Let us first write down a simple model for thin film electrons. In the XY plane, we apply periodic boundary conditions. The electron confinement in the z-direction is usually characterized by an infinite well potential with its boundaries at $z=0$ and $z=d$. However this model cannot permit tunneling of thin film electrons to the TI side since the amplitude of the electron wavefunction is zero at the interface. To allow for tunneling, a simple way is to impose open boundary conditions at the interface so that the amplitude is maximum at the interface. As a result, the momentum in the z-direction gets quantized as $k_z = (n - 1/2)\pi / d$ where $n = 1, 2,..$, and the z-dependence of the electron wavefunction becomes $\psi_n(z) = \sqrt{2/d}\cos\left((n - 1/2)\pi z/ d\right)$. Thus, the Hamiltonian governing the dynamics of thin film electrons deposited over the TI has the form,
\begin{eqnarray}
    \mathcal{H}^{\text{tf}} =&&  \sum_{n, s}\int \frac{d^{2}\textbf{k}}{(2\pi)^2} c^{\dagger}_{\textbf{k}, n} h^{\text{tf}}_{\textbf{k}, n} c_{\textbf{k}, n}\label{HQW}
    \end{eqnarray}
    where
    \begin{eqnarray}
          h^{\text{tf}}_{\textbf{k}, n} &=& \epsilon^{\text{tf}}_{\textbf{k},n}\hat{I} \\ &=& \biggl[\frac{\hbar^{2}k^{2}}{2m^{*}} + (n - 1/2)^2\frac{\pi^{2}\hbar^2}{2m^{*}d^2} \biggr]\hat{I}\nonumber  
    \end{eqnarray}
where  $c_{\textbf{k}, n} = \begin{bmatrix} c_{\textbf{k}, n, \uparrow} & c_{\textbf{k}, n, \downarrow} \end{bmatrix}$ is the  creation operator for an electron at the $n$th quantum well state with in-plane momentum $\textbf{k} = (k_x, k_y)$ in the thin film. $\hat{I}$ is just an identity matrix to emphasize that $h^{\text{tf}}$ is a $2\times 2$ matrix in the spin-1/2 space.

The effective Hamiltonian that describes the surface states of a topological insulator is,
\begin{eqnarray}
    \mathcal{H}^{\text{surf}} &=& \int \frac{d^{2}\textbf{k}}{(2\pi)^2}\chi^{\dagger}_{\textbf{k}} h^{\text{surf}}_{\textbf{k}} \chi_{\textbf{k}}\label{HTI}
\end{eqnarray}
where
\begin{eqnarray}
     h^{\text{surf}}_{\textbf{k}} = A_0\left(s_x k_y - s_y k_x\right) + E_{0}\nonumber
\end{eqnarray}
where $\chi^\dagger = \begin{bmatrix} \chi^\dagger_\uparrow & \chi^\dagger_\downarrow \end{bmatrix} $ is the creation operator of the surface electron. $s_x, s_y$ are Pauli matrices in the spin-1/2 space. $A_0$ describes the strength of spin-orbit coupling. $E_0$ is the energy at the Dirac point. Due to the presence of spin-orbit coupling, the Hamiltonian does not have spin-rotation symmetry. Rather it is diagonal in the helicity basis. The energy eigenstates in the helicity basis are given by,
\begin{eqnarray}
    \epsilon^{\text{surf}}_{\textbf{k},\pm} = \pm A_0 |\textbf{k}| + E_{0}\label{HTI2}
\end{eqnarray}

Assuming that the tunneling process is spin-independent, the simplest model that can describe the hybridization of the surface states of the TI with the quantum well states of the thin film is given by,
\begin{eqnarray}
    \mathcal{H}^t = t\int d^2\textbf{r}\, \left(\chi^\dagger(\textbf{r}) \Psi(\textbf{r}, z = 0) + h.c\right)  
\end{eqnarray}
here $\chi^\dagger(\textbf{r})$ is the spinor field operator that creates a topological surface electron at in-plane position $\textbf{r} = (x,y)$. $\Psi(\textbf{r})$ is the spinor field operator thin film electrons with open boundary conditions. In the k-space, the Hamiltonian is of the form,
\begin{eqnarray}
    \mathcal{H}^t &=& t_d \,\sum_{n}\int \frac{d^{2}\textbf{k}}{(2\pi)^2} \chi^{\dagger}_{\textbf{k}} c_{\textbf{k}, n} + h.c  \label{Ht} \\ t_d &=& \frac{t}{\sqrt{d}} \nonumber
\end{eqnarray}

We find that the effective tunneling strength scales as a function of the thin film thickness $d$ as a result of quantum confinement in the $z$-direction. The surface area in the $xy$-plane given by $L_xL_y$ is set to unity throughout this paper.

 In this article, we ignore the possibility of multi-band tunneling. This is a good approximation provided we work in the limit where the energy difference between successive thin film QW bands is greater than the bulk energy gap of the topological insulator. In this limit, there is effectively only one QW band on which the tunneling effects due to TI surface electrons are significant. Since the chemical potential is aligned within the bulk energy gap of the TI, this QW band will be the topmost conduction band of the thin film. In other words, this band will be the one closest to the Dirac point of the TI surface in terms of energy. Tunneling effects on other QW bands are perturbative which is not the focus of our study in this section. Quantitatively, the effective model that we introduce in this article works well only when the following condition is satisfied,
 \begin{eqnarray}
     |\epsilon^{\text{tf}}_{\textbf{k}=0,N} - \epsilon^{\text{tf}}_{\textbf{k}=0,N\pm 1}| \geq m \label{condition}
 \end{eqnarray}
 where $m$ is the mass gap of the topological insulator and $N$ is the band index of the thin film QW band that is energetically closest to the Dirac point of the TI surface. Once this condition is satisfied, we can conveniently ignore the tunneling effects on all other $n \neq N$ bands. This setup is illustrated schematically in Fig.\ref{TI-thfmschematic}(b). Then the simplified effective Hamiltonian of the electronic states involved in tunneling becomes,   
\begin{eqnarray}
   \mathcal{H}^{\text{hbd}} =&& \int \frac{d^{2}\textbf{k}}{(2\pi)^2} \biggl[ c^{\dagger}_{\textbf{k},N}h^{\text{tf}}_{\textbf{k}, N}c_{\textbf{k}, N} + \chi^{\dagger}_{\textbf{k}} h^{\text{surf}}_{\textbf{k}} \chi_{\textbf{k}}\nonumber\\ +&& t_d \,c^{\dagger}_{\textbf{k},N}\chi_{\textbf{k}} + \text{h.c} \biggr] \label{band}
\end{eqnarray}

\subsubsection{Hybridization at the interface}
Turning on $t$ results in thin film electrons tunneling to the TI surface side and vice versa. Tunneling effects will be significant when $\frac{t}{|\triangle E|} > 1$, where $\triangle E $ is the difference in energy between the initial and the final state. In this case, a perturbative treatment won't be sufficient. Here we shall understand the effects of tunneling in a non-perturbative manner. The full Hamiltonian is diagonalized exactly and the properties of the resulting hybrid electrons are studied. To diagonalize the Hamiltonian, we shall define a $SU(2)$ space $\sigma_i(i = x,y,z)$ to model the spatial profile of the electrons. In this space, the single-particle Hamiltonian in the momentum space becomes,
\begin{eqnarray}
    h^{\text{hbd}}_{\textbf{k}} &=& \left(
\begin{array}{ccc}
h^{\text{tf}}_{\textbf{k}, N} & t_d \\ t_d & h^{\text{surf}}_{\textbf{k}}
\end{array}\right)\nonumber\\ &=& I \otimes M_{\textbf{k}, N} + \sigma_z \otimes \delta_{\textbf{k}, N}  +  \sigma_x \otimes I t_d
\end{eqnarray}\
in the basis $\Gamma_{\textbf{k},N}^{\dagger} = \left(\begin{array}{cccc}c^{\dagger}_{\textbf{k}, N, \uparrow} & c^{\dagger}_{\textbf{k}, N, \downarrow} & \chi^\dagger_{\textbf{k},\uparrow} & \chi^\dagger_{\textbf{k},\downarrow}  \end{array}\right)$.
Here $\delta_{\textbf{k}, N}$ and $M_{\textbf{k}, N}$ are $2\times2$ matrices in the spin-1/2 space with the respective definitions:
\begin{eqnarray}
    \delta_{\textbf{k}, N} &=& \left( h^{\text{tf}}_{\textbf{k}, N} - h^{\text{surf}}_{\textbf{k}}\right)/2\nonumber \\ 
M_{\textbf{k}, N} &=& \left(h^{\text{tf}}_{\textbf{k}, N} + h^{\text{surf}}_{\textbf{k}}\right)/2 \label{deltaM}
\end{eqnarray}
Since we discuss the hybridization effect only on the $N$th band in the thin film, the index $N$ will be dropped from now on. But do note that the unitary matrix elements do depend on the value of $N$ which in turn is connected to the thickness of the thin film.
The Hamiltonian can be diagonalized in the $\sigma$ space by performing a unitary transformation with the following unitary matrix,
\begin{eqnarray}
   U_{\textbf{k}} = \left(\begin{array}{ccc}     \cos{\frac{\theta_{\textbf{k}}}{2}} & \sin{\frac{\theta_{\textbf{k}}}{2}} \\ -\sin{\frac{\theta_{\textbf{k}}}{2}} & \cos{\frac{\theta_{\textbf{k}}}{2}}
   \end{array}\right), \,\, \cos \theta_{\textbf{k}} = \frac{\delta_{\textbf{k}}}{\sqrt{\delta^2_{\textbf{k}} + \frac{t^2}{d}}}\label{unitary}
\end{eqnarray}
The Hamiltonian after rotation attains the following diagonal form,
\begin{eqnarray}
    \mathcal{H}^{\text{hbd}} = \int \frac{d^{2}\textbf{k}}{(2\pi)^2} \left[d^{\dagger}_{\textbf{k},t}h_{\textbf{k},t} d_{\textbf{k},t} + d^{\dagger}_{\textbf{k},b}h_{\textbf{k},b} d_{\textbf{k},b} \right]
\end{eqnarray}
where $d^{\dagger}_{\textbf{k},t(b)} = \begin{bmatrix}
    d^{\dagger}_{\textbf{k},t(b),\uparrow} & d^{\dagger}_{\textbf{k},t(b),\downarrow}
\end{bmatrix}$ are two-component spinors in the spin basis. 
 $h_{\textbf{k},t(b)}$ have the following definitions,
 \begin{subequations}
\begin{eqnarray}
h_{\textbf{k},t} = M_{\textbf{k}} + \sqrt{\delta^2_{\textbf{k}} + t^2_d}\\h_{\textbf{k},b} = M_{\textbf{k}} - \sqrt{\delta^2_{\textbf{k}} + t^2_d}    
\end{eqnarray}     
 \end{subequations}
 \begin{figure}[b]
\includegraphics[width=8.50cm, height=3.03cm]{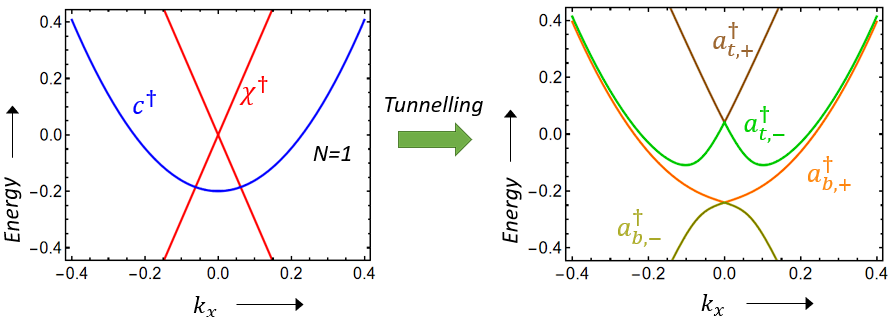}
\caption{Energy spectrum of the $N$th band of the thin film and the surface Dirac cone before and after the tunneling is turned on. Here $N$ is set to unity. The definitions of the creation operators shown in the picture are contained in the main text.(See Eqns.\ref{HQW}, \ref{HTI} and \ref{Hhbd}).}
\label{spctrm}
\end{figure}

here the index $t$ and $b$ represent the 'top' and 'bottom' bands respectively. This splitting is a result of the tunneling of single-particle states between the two sides of the hybrid. In addition, we find that  $h_{t(b)}$ are $2\times2$ matrices in the spin-1/2 space. $h_{\textbf{k},t(b)}$  has terms proportional to $s_x k_y - s_y k_x$ implying that the hybrid states acquired an emergent spin-orbit coupling. Tunneling essentially resulted in hybridizing the thin film QW state and the TI surface state. Due to this induced helical spin structure of the hybrid states, it is better to write the full Hamiltonian on a helicity basis. We define the following set of creation operators,

\begin{eqnarray}
    d^{\dagger}_{\textbf{k},t(b)} = a^{\dagger}_{\textbf{k},t(b)}\Pi^{\dagger}_{\textbf{k}}, && \Pi_{\textbf{k}} = \frac{1}{\sqrt{2}}\left(\begin{array}{cc}
        1 & 1 \\
        e^{i\phi_{\textbf{k}}} & -e^{i\phi_{\textbf{k}}}
    \end{array} \right)\nonumber\\ e^{i\phi_{k}} &=& \frac{k_y - i k_{x}}{|\textbf{k}|}  \label{helicitybasisdef}
\end{eqnarray}

where $a^{\dagger}_{\textbf{k},t(b)} = \begin{bmatrix}
    a^{\dagger}_{\textbf{k},t(b),+} & a^{\dagger}_{\textbf{k},t(b),-}
\end{bmatrix}$. Here $(+)$ and $(-)$ represent states with positive and negative helicity respectively. In this helicity basis, the single-particle Hamiltonian  has the following diagonal representation,
\begin{subequations}
\begin{eqnarray}
    \mathcal{H}^{\text{hbd}} &=& \int \frac{d^{2}\textbf{k}}{(2\pi)^2} \biggl[a^{\dagger}_{\textbf{k},t, +}\epsilon^{\text{hbd}}_{\textbf{k},t,+}a_{\textbf{k},t, +}\nonumber\\ &+& a^{\dagger}_{\textbf{k},t, -}\epsilon^{\text{hbd}}_{\textbf{k},t,-}a_{\textbf{k},t,-} + a^{\dagger}_{\textbf{k},b, +}\epsilon^{\text{hbd}}_{\textbf{k},+}a_{\textbf{k},b, +}\nonumber\\ &+& a^{\dagger}_{\textbf{k},b, -}\epsilon^{\text{hbd}}_{\textbf{k},b,-}a_{\textbf{k},b,-} \biggr]\\ \epsilon^{\text{hbd}}_{\textbf{k},t,\pm} &=& \frac{\epsilon^{\text{tf}}_{\textbf{k},N} + \epsilon^{\text{surf}}_{\textbf{k},\pm}}{2} + \sqrt{\left( \frac{\epsilon^{\text{tf}}_{\textbf{k},N} - \epsilon^{\text{surf}}_{\textbf{k},\pm}}{2}\right)^{2} + t^2_d}\nonumber\\ \\ \epsilon^{\text{hbd}}_{\textbf{k},b,\pm} &=& \frac{\epsilon^{\text{tf}}_{\textbf{k},N} + \epsilon^{\text{surf}}_{\textbf{k},\pm}}{2} - \sqrt{\left( \frac{\epsilon^{\text{tf}}_{\textbf{k},N} - \epsilon^{\text{surf}}_{\textbf{k},\pm}}{2}\right)^{2} + t^2_d}\nonumber \\ 
\end{eqnarray}    
\label{Hhbd}
\end{subequations}
 Fig.\ref{spctrm} shows an example of the energy spectrum before and after the tunneling.  
 Given that the condition in Eqn.\ref{condition} is satisfied, the tunneling effect on the thin QW bands of index $n \neq N$ is perturbative and hence they are ignored. Therefore, the single-particle Hamiltonian of all these $n\neq N$ QW bands is unaffected by the tunneling and retains the form given in Eqn.\ref{HQW}. The electronic states in these bands will play huge role in the pairing physics especially in the large-$N$ limit, as we shall see later in this article.

\section{\label{sec:level3}Effective Pairing Hamiltonian}
We examined the physics of single-particle tunneling at the thin film-TI hybrid in the preceding section. We discovered that non-perturbative tunneling results in the hybridization of the surface bands with the thin film's resonant quantum-well band. We now have a four-band model with single particle states that are a linear superposition of the thin film state and the surface state. As a result, it is possible that the hybrid states couple with the phonons in the thin film. The effective short-ranged pairing Hamiltonian that explains the interactions of the hybrid electrons with one another and with the thin film electrons belonging to the inner bands is derived in this section starting with the fundamental electron-phonon coupling Hamiltonian.
\subsection{Phonon-mediated interaction potential between thin film electrons}
\subsubsection{2D electron-phonon coupling Hamiltonian}
Similar to electrons, phonons in the thin film are also spatially confined within the range $z=0$ and $z=d$. As a result, the phonon spectrum also gets quantized resulting in the formation of 2D QW bands indexed by the integer $l$. We implement open boundary conditions at the thin film-TI interface. The phonon spectrum becomes, $E_{\text{ph}}(\textbf{q}, l) = \hbar c \sqrt{q^{2} + \left(\frac{(l-1/2)\pi}{d}\right)^2}$, where $l$ is an integer identifying the confined slab phonon mode. The electron-phonon coupling Hamiltonian in 3D has the form,
\begin{eqnarray}
   \mathcal{H}_{\text{e-ph}} = G_{\text{fp}} \int d^{2}\textbf{r}dz\, \Psi^{\dagger}(\textbf{r},z)\Vec{\nabla}.\Vec{\Phi}(\textbf{r},z)\Psi(\textbf{r},z) 
\end{eqnarray}
where $\Psi(\textbf{r})$ is the 2-component electron field operator  and $\Phi_{i}(\textbf{r})(i=x,y,z)$ is the phonon field operator in the thin film with the following definitions,
\begin{eqnarray}
     \Psi(\textbf{r},z) &=& \sum_{n}\int \frac{d^{2}\textbf{k}}{(2\pi)^{2}} \, \psi_{n}(z) e^{i\textbf{k}.\textbf{r}} c_{\textbf{k}, n}\nonumber \\
     \Phi_{i}(\textbf{r},z) &=& \sum_{l}\int \frac{d^{2}\textbf{q}}{(2\pi)^{2}}\phi_l(z)\frac{e^{i\textbf{q}.\textbf{r}}}{2\sqrt{E_{ph}(\textbf{q}, l)}} \left[b_{\textbf{q}, l, i} + b^{\dagger}_{-\textbf{q}, l, i} \right]\nonumber\\
\end{eqnarray}
where $\psi_n(z) = \sqrt{\frac{2}{d}} \cos \left(\frac{(n-1/2)\pi z}{d}\right)$ and $\phi_l(z) =  \sqrt{\frac{2}{d}} \cos \left(\frac{(l-1/2)\pi z}{d}\right)$. Integrating out the z-degrees of freedom, we obtain the following effective 2D Hamiltonian, 
\begin{equation}
    \mathcal{H}_{\text{e-ph}} = \sum_{n, n', l} g^{l}_{n,n^{'}}(d) \int d^{2}\textbf{r}\,\Psi^{\dagger}_{n'}(\textbf{r})\vec{\nabla}.\vec{\Phi}_{l}(\textbf{r})\Psi_{n}(\textbf{r})\label{eph}
\end{equation}
Here $\Psi_{n}$ is the effective 2D electron field operator for an electron with band index $n$. Similar definition holds for $\vec{\Phi}_{l}$. The scattering matrix $g^{l}_{n,n^{'}}(d)$ is given by
\begin{eqnarray}
    g^{l}_{n,n^{'}}(d) &=& (-1)^{n+n'-l}\frac{G_{\text{fp}}}{\pi}\sqrt{\frac{2}{d}}\biggl[\frac{l - \frac{1}{2}}{( l - \frac{1}{2})^2 - (n - n')^2}\nonumber \\ &-& \frac{l - \frac{1}{2}}{( l - \frac{1}{2})^2 - (n + n' - 1)^2}\biggr]
\end{eqnarray}    
We find here that coupling with phonons can lead to interband scattering of electrons in the thin film.


\subsubsection{Pairing potential matrix}
 It is well known that coupling with phonons leads to an effective electron-electron interaction that could be attractive under certain conditions. The minimal BCS pairing Hamiltonian that emerges out of the coupling term in Eqn.\ref{eph}, has the following form,
 \begin{eqnarray}
      \mathcal{H}_{I} &=& \sum_{n,n'}\mathcal{H}_{I}(n, n')\nonumber\\ &=& - \sum_{n,n'}\int \frac{d^{2}\textbf{k}}{(2\pi)^{2}}\frac{d^{2}\textbf{p}}{(2\pi)^{2}}\nonumber \\ && V^{n,n'}_{\textbf{k},\textbf{p}}c^{\dagger}_{\textbf{p}, n} s_y c^{\dagger T}_{-\textbf{p}, n} c^{T}_{-\textbf{k},n'}s_y c_{\textbf{k}, n'}\label{H_I} 
 \end{eqnarray}
where the pairing potential $V_{n,n'}$ has the form,
\begin{eqnarray}
 V^{n,n'}_{\textbf{k},\textbf{p}} = \begin{cases} \sum\limits_{l=1}^{l_{max}} |g^{l}_{n,n'}(d)|^2,& - \omega_{D} < \xi_{\textbf{k},n},\xi_{\textbf{p},n'} < \omega_{D}\nonumber  \\ 
  0,& \text{else}
  \end{cases} \\ \label{V}
\end{eqnarray}
 Here $\omega_{D}$ is the Debye frequency of the thin film. $\xi_{\textbf{k},n}$ is the single-particle energy of the thin film electrons measured from the chemical potential. The electron-phonon coupling matrix $g_{n,n'}$ is summed over all the slab phonon modes up to $l_{max}$. It is the maximum value that a phonon mode could have in the thin film at a given thickness $'d'$. To find its value, recall that Debye frequency sets the UV cut-off for the energy of lattice vibrations. Hence $l_{max}$ can be calculated by taking the integer part of the expression $d \left(k_{D}/\pi\right)$, where $k_{D}$ is the Debye momentum. A comprehensive study of the thin film superconductivity with attractive interaction mediated by confined phonons was conducted in ref.\cite{Sarma(2000)}.

 An important consequence of the dimensional reduction applied in the context of interactions\cite{saran(2022)} is that the effective 2D interaction potential  acquires a scaling dependence on the thin film thickness as,
 \begin{equation}
     V^{n,n'}_{\textbf{k},\textbf{p}} \propto \frac{1}{d}
 \end{equation}
  Thus the attractive interaction increases with reducing thickness. This implies that the attractive interaction is maximum in the ultrathin($N=1$) limit of the thin film. We shall use this scaling relation in the later part of this paper in order to enhance the attractive interaction between surface fermions.

\subsection{The general interaction Hamiltonian of the thin film-TI hybrid}

When the tunneling is turned on, the thin film band which is close to the Dirac point of the TI surface is hybridized. Let $N$ be the index of the band that is hybridized. As mentioned before, we consider only the limit when the $N\pm1$ bands are separated from the $N$th band by a magnitude of at least the order of bulk energy gap of the TI (See Eqn.\ref{condition}). So, the effects of hybridization on all these $n \neq N$ bands are ignored. Now coming back to the $N$th band, hybridization with the surface Dirac cone implies that the electronic states in that QW band are no longer diagonal in the thin film basis. The hybrid states are in a linear superposition of the thin film and the TI surface states. The emergent excitations of this hybrid system are the states  $d^{\dagger}_{\textbf{k},t(b)}\ket{0}$ in the spin basis. It is even easier to study the interaction if we could rotate the states to the helicity basis since the hybrid states are diagonal in the helicity basis. So we project the interaction Hamiltonian $\mathcal{H}_I$ of the resonant band indexed by $N$ into the basis spanned by $a_{\textbf{k},t(b),\pm}$ states (defined in Eqn.\ref{helicitybasisdef}). 

After the projection, the full Hamiltonian $\mathcal{H}_{I}$ can be divided into essentially three terms. The first term is the Hamiltonian describing the attractive interaction between the helical hybrid fermions. Secondly, we have the term describing attractive interaction between the hybridized fermions and the trivial fermions of all the $n\neq N$ thin film transverse bands. Lastly, we have the interaction Hamiltonian for the fermions in the thin film unaffected by hybridization. In doing this projection, terms that describe interband pairing between the helical fermions have been ignored. This is a good approximation in the BCS limit. We shall write down the three terms in the Hamiltonian explicitly below,
\begin{eqnarray}
    \mathcal{H}_{I} = \mathcal{H}^{\text{hbd-hbd}}_{I} + \mathcal{H}^{\text{hbd-tf}}_{I} + \mathcal{H}^{\text{tf-tf}}_{I}\label{HIfull}
\end{eqnarray}
Now we shall derive these three terms in the Hamiltonian starting from the fundamental s-wave pairing Hamiltonian in the thin film. The details of the derivation are given in the appendix \ref{deriv:HI}. 
\subsubsection{Hamiltonian for Interaction between hybrid fermions ($\mathcal{H}^{\text{hbd}-\text{hbd}}$) }
Here we shall derive the pairing Hamiltonian that describes the attractive interaction between the helical hybrid fermions. Before the tunneling was switched on, the interaction between electrons in the $N$th band of the thin film is described by the following Hamiltonian,
\begin{eqnarray}
          \mathcal{H}_{I}(N, N) &=& - \int \frac{d^{2}\textbf{k}}{(2\pi)^{2}}\frac{d^{2}\textbf{p}}{(2\pi)^{2}}\nonumber\\ && \,V^{N,N}_{\textbf{k},\textbf{p}}c^{\dagger}_{\textbf{p}, N} s_y c^{\dagger T}_{-\textbf{p}, N} c^{T}_{-\textbf{k},N}s_y c_{\textbf{k}, N}
\end{eqnarray}
Once the tunneling is switched on, the electronic states in the $N$th band are hybridized and we have a 4-band model with a helical spin texture. So, it is better that the interaction Hamiltonian be written down in the helicity basis. Before we write down the Hamiltonian, we shall define the notations used to identify all four hybrid bands. Let $m$, $m'$ run over the band indices $t$(top) and $b$(bottom). Similarly, $\lambda$ and $\lambda'$ run over the $+$ and $-$ helicity branches. Using this set of indices, we can write down the following interaction Hamiltonian that describes all possible pairing interactions(except the inter-band pairing) between the four hybrid bands:
\begin{subequations}
\begin{multline}
    \mathcal{H}^{\text{hbd-hbd}}_{I} = - \sum_{\alpha, \beta}\int \frac{d^{2}\textbf{k}}{(2\pi)^{2}}\frac{d^{2}\textbf{p}}{(2\pi)^{2}} e^{i\left(\phi_{\textbf{p}} - \phi_{\textbf{k}} \right)} \\ \lambda\lambda'J^{\alpha,\beta}_{\textbf{k},\textbf{p}}a^{\dagger}_{\textbf{k},\alpha}a^{\dagger}_{-\textbf{k},\alpha}a_{-\textbf{p},\beta}a_{\textbf{p},\beta}\label{HI1}
\end{multline}
\begin{eqnarray}    J^{\alpha,\beta}_{\textbf{k},\textbf{p}} = V^{N,N}_{\textbf{k},\textbf{p}}Z^{\alpha}_{\textbf{k}}Z^{\beta}_{\textbf{p}} \label{J_gen}
\end{eqnarray}
\end{subequations}
 Here $\alpha = (m,\lambda) \,\, \beta = (m', \lambda')$ is used as a shorthand notation to denote the band indices. Note that $\lambda\lambda' = -1$ if the scattering is between bands of opposite helicity. 
 
 $Z^{\alpha}_{\textbf{k}}$ can be identified as the wavefunction renormalization of a hybridized electronic state as a result of tunneling with respect to a thin film state without tunneling. This implies that $Z^{\alpha}_{\textbf{k}} = 1$ for a thin film state and $Z^{\alpha}_{\textbf{k}} = 0$ for TI surface state before the tunneling was turned on. They have the following structure,
    \begin{eqnarray}
    Z^{(t,\pm)}_{\textbf{k}} &=& \frac{1}{2}\left( 1 + \frac{\delta_{\textbf{k},\pm}}{\sqrt{\delta^2_{\textbf{k},\pm} + \frac{t^2}{d}}}\right)\label{Z1}\nonumber\\ Z^{(b,\pm)}_{\textbf{k}} &=& \frac{1}{2}\left( 1 - \frac{\delta_{\textbf{k},\pm}}{\sqrt{\delta^2_{\textbf{k},\pm} + \frac{t^2}{d}}}\right)\label{Z2}\nonumber\\ \delta_{\textbf{k},\pm} &=& \frac{1}{2}\left(\epsilon^{\text{tf}}_{\textbf{k},N} - \epsilon^{\text{surf}}_{\textbf{k},\pm} \right)\nonumber \\ \label{Zfactor}
\end{eqnarray}
 So we find here that, as a result of tunneling, a pairing potential exists between the helical hybrid fermions and it is proportional to the square of the renormalization factors of the bands corresponding to the initial and final states of the Kramers pair of electrons involved in pairing. This makes physical sense because the $Z$-factor determines the probability that an electron is in the thin film side of the interface. Only the electrons in the thin film side of the interface will experience an attractive interaction mediated by phonons. If $Z^{\alpha}_{\textbf{k}} = 1$ for an electronic state of momentum $\textbf{k}$ and in a hybrid band indexed by $\alpha$, the electronic state is completely in the thin film side of the interface and experience the full attractive interaction. But in this case, the electronic state will not have the helical spin texture induced by the TI surface. On the other hand, if $Z^{\alpha}_{\textbf{k}} = 0$ for an electronic state in the hybrid band, then the electron is entirely on the TI side of the interface and does not experience an attractive interaction. So we have to fine-tune the material parameters such that both the effects, the helical spin texture, and the attractive interaction are substantial. We shall show in this article quantitatively that this can be achieved by fine-tuning the thickness to 'quantum well resonance' at the Dirac point. A detailed discussion of this phenomenon will be presented in the next section.

\subsubsection{Interaction between hybrid fermions and the thin film fermions in the $n \neq N$ band  $\mathcal{H}^{\text{hbd-tf}}$}
In the limit that we are working, hybridization effects are substantial only for the thin film QW band at $n=N$. All the other $n\neq N$ bands are much above or much below the Dirac point of the TI surface so that the tunneling effects due to surface fermions are negligible. But it is possible that the hybrid fermions can still experience attractive interaction with the thin film electrons lying in all of the $n\neq N$ bands. This effect is captured by the interband scattering terms of the thin film interaction Hamiltonian given in Eqn.\ref{H_I}. Before tunneling is introduced, it is possible that a singlet Cooper pair of electrons in the $N$th band can scatter to any of the $n\neq N$ bands. The Hamiltonian describing such a process can be read out from the full interaction Hamiltonian given in Eqn.\ref{H_I} by fixing $n'$ to $N$ and letting $n$ run over all $n\neq N$.
\begin{eqnarray}
          \sum_{n\neq N}\mathcal{H}_{I}(N, n) = - \sum_{n\neq N}\int \frac{d^{2}\textbf{k}}{(2\pi)^{2}}\frac{d^{2}\textbf{p}}{(2\pi)^{2}}\nonumber \\ \,V^{N,N}_{\textbf{k},\textbf{p}}c^{\dagger}_{\textbf{p}, n} s_y c^{\dagger T}_{-\textbf{p}, n} c^{T}_{-\textbf{k},N}s_y c_{\textbf{k}, N}    
\end{eqnarray}
Once the tunneling is switched on, the Cooper pair $c^{T}_{-\textbf{k},N}s_y c_{\textbf{k}, N}$ is projected to the helicity basis of the $t$ and $b$ hybrid bands. In doing this, we arrive at an interaction Hamiltonian that describes the attractive interaction between the hybrid fermions and the off-resonance thin film fermions. Let us call the Hamiltonian by the name $\mathcal{H}^{\text{hbd-tf}}_{I}$ and has the following definition,
\begin{subequations}
\begin{eqnarray}
          \mathcal{H}^{\text{hbd-tf}}_{I} = - \sum_{n\neq N}\sum_{\alpha}\int \frac{d^{2}\textbf{k}}{(2\pi)^{2}}\frac{d^{2}\textbf{p}}{(2\pi)^{2}}e^{i\phi_{\textbf{p}}}\nonumber \\ \,\lambda K^{n,\alpha}_{\textbf{k},\textbf{p}} c^{\dagger}_{\textbf{k}, n}(-is_y) c^{\dagger T}_{-\textbf{k}, n} a_{-\textbf{p},\alpha} a_{\textbf{p},\alpha}\label{HI2}    
\end{eqnarray}
\begin{eqnarray}    K^{n,\alpha}_{\textbf{k},\textbf{p}} = V^{n,N}_{\textbf{k},\textbf{p}}Z^{n}_{\textbf{k}}Z^{\alpha}_{\textbf{p}}
\end{eqnarray}
\end{subequations} 
Note that $Z^{n}_{\textbf{k}} = 1$ for all $k$ and $n\neq N$ since it corresponds to the renormalization factor of the thin film electrons which did not participate in tunneling. It has been included in the expression only for the purpose of generality. So here we find that even though the thin film electrons in the $n \neq N$ bands do not participate in tunneling, they do contribute to the superconducting phase of the hybrid fermions.

\subsubsection{Interaction between all the $n \neq N$ band thin film fermions ($\mathcal{H}^{\text{tf-tf}}_{I}$)}
It is also important to consider the attractive interaction between the electrons in the $n\neq N$ bands that were not part of the tunneling. It is just the trivial BCS singlet pairing Hamiltonian. It is found by summing over $\mathcal{H}_{I}(n, n')$ defined in Eqn.\ref{H_I} for all $n,n' \neq N$. Let us call this Hamiltonian as $\mathcal{H}^{\text{tf-tf}}_{I}$. It has the form,
 \begin{eqnarray}
      \mathcal{H}^{\text{tf-tf}}_{I} &=& - \sum_{n,n' \neq N}\int \frac{d^{2}\textbf{k}}{(2\pi)^{2}}\frac{d^{2}\textbf{p}}{(2\pi)^{2}}\nonumber \\ && V^{n,n'}_{\textbf{k},\textbf{p}}c^{\dagger}_{\textbf{p}, n} s_y c^{\dagger T}_{-\textbf{p}, n} c^{T}_{-\textbf{k},n'}s_y c_{\textbf{k}, n'} \label{HI3} 
 \end{eqnarray}
The full interaction Hamiltonian of the TI-thin film hybrid is now the sum of all three terms as given in Eqn.\ref{HIfull}.
\section{\label{sec:level4}Z-factor and the quantum-well resonance}

In section II, we studied the single-particle tunneling of electronic states in the topological surface to the QW thin film band lying closest to it. The tunneling effectively results in the hybridization of the electronic states and leads to the formation of four spin-split hybrid bands, with an emergent helical spin texture for each of them.

In section III, we found that these helical hybrid electrons can couple with the confined phonons of the thin film and could result in an effective attractive interaction between them. The effect of tunneling is taken into account in the interaction strength by the renormalization factor $Z^{\alpha}_{\textbf{k}}$ defined in Eqns.\ref{Zfactor}. For instance,  one can show that the type of pairing between two electrons with renormalization factors equal to unity will be trivial s-wave-like. This is because these electrons lie entirely in the thin film side and the tunneling effect on them is negligible. The other extreme is when the renormalization factor of the electrons is zero. This corresponds to the non-interacting surface electrons.

From these intuitive arguments, one can anticipate that the ideal choice for the renormalization factor of an electronic state will be $1/2$.  It is at this limit the tunneling effect is maximum. This implies that the surface states that are initially non-interacting will acquire maximum attractive interaction in this limit. This is because it is the tunneling that actually induces an effective attractive interaction between surface fermions. In order to realize this maximum tunneling effect, the corresponding electronic states on both sides of the interface must be degenerate. In other words, the electronic states should be in quantum-well resonance. In this section, we will show this explicitly by studying the behavior of the renormalization factor as a function of the detuning parameter defined at the Dirac point. 

The renormalization factors were defined in Eqn.\ref{Zfactor} as a function of the band indices and momentum $\textbf{k}$. Since there are four hybrid bands, we have four renormalization factors for a fixed momentum $\textbf{k}$. One can show that they follow a general relationship,
\begin{eqnarray}
    Z^{(t,+)}_{\textbf{k}} + Z^{(b,+)}_{\textbf{k}} &=& 1\nonumber  \\ Z^{(t,-)}_{\textbf{k}} + Z^{(b,-)}_{\textbf{k}} &=& 1 
\end{eqnarray}
for any momentum state $\textbf{k}$. This implies that for a fixed helicity if one hybrid band is on the thin film side, the other band lies on the TI surface side. We are mostly interested in the interacting dynamics of the electronic states near the Dirac point. Therefore we set the momentum $k = 0$ in the above equations and study the evolution of the $Z$-factors as a function of the detuning parameter also defined at zero momentum. At $k=0$, there is a further simplification. We find that due to the crossing of the two helicity branches at the Dirac point, the respective $Z$ factors turn out to be equal. That is,
\begin{figure}[b]
\includegraphics[width=5.5cm, height=3.7cm]{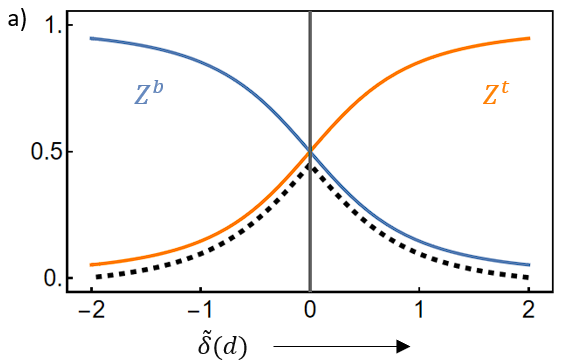}
\bigbreak
\includegraphics[width=8.5cm, height=5.5cm]{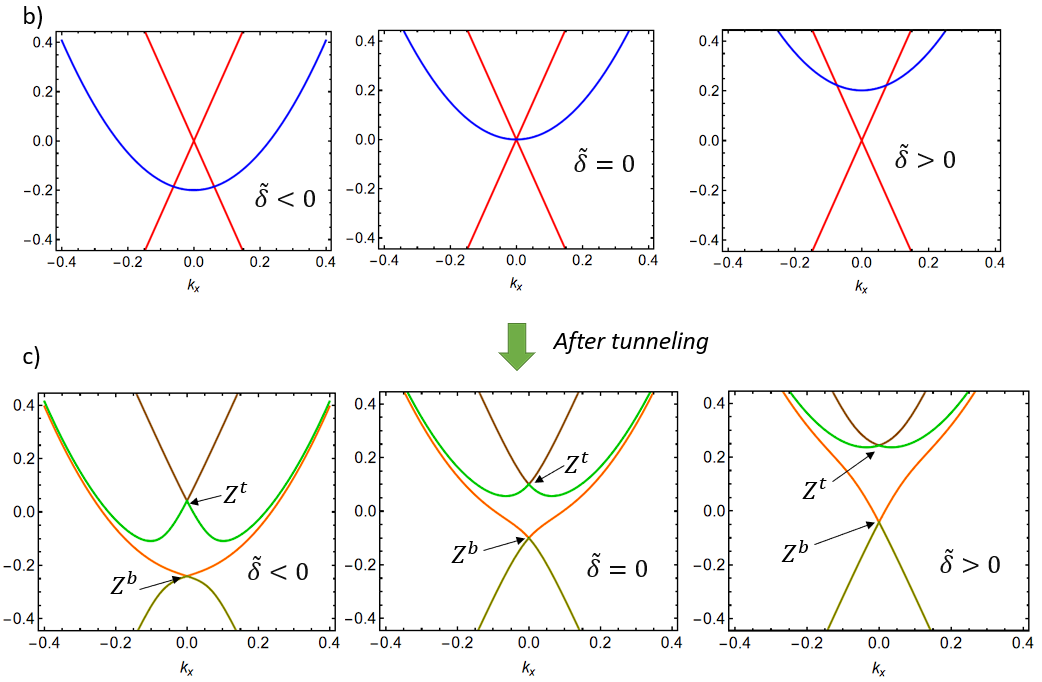}
\caption{\label{qwsproof} a) Plot showing the evolution of the wavefunction renormalization Z-factors defined at the Dirac point, $Z^{t}$ and $Z^b$(defined in Eqns.\ref{Zfactor},\ref{ZtZb}) corresponding to the top and the bottom hybrid bands respectively as a function of the detuning parameter $\tilde{\delta}$. The black dashed lines trace the Z-factor of the surface fermions. The Z-factor and hence the surface attractive interaction that is proportional to $Z^2$ (Eqns.\ref{J_gen}) is maximized at resonance($\tilde{\delta} = 0$). Also shown are the energy spectra of the thin film QW band and the surface Dirac cone at three different values of the detuning parameter before (b) and after (c) the tunneling is turned on. The arrowheads show the respective Dirac points at which we calculated $Z^t$ and $Z^b$ of the top and bottom bands. The detuning parameter is tuned by fine-tuning the thin film thickness.}
\end{figure}
\begin{eqnarray}
    Z^{(t,+)}_{\textbf{k} = 0} = Z^{(t,-)}_{\textbf{k} = 0}, \,\,\text{and} \,\, Z^{(b,+)}_{\textbf{k} = 0} = Z^{(b,-)}_{\textbf{k} = 0}
\end{eqnarray}
So at $k=0$, we essentially have ended up with just two $Z$ factors subject to the constraint that their sum must be equal to unity. We shall make the following redefinitions,
\begin{eqnarray}
    Z^{t} = Z^{(t,\pm)}_{\textbf{k} = 0}\,\,&&\text{and} \,\,Z^{b} = Z^{(b,\pm)}_{\textbf{k} = 0}\label{ZtZb} \\ \text{so we have}\,\,Z^t + Z^b &=& 1\nonumber 
\end{eqnarray}
Now we shall define the detuning parameter at $k = 0$. It has the form,
\begin{eqnarray}
    \tilde{\delta}(d) = \frac{\delta_{\textbf{k}=0,N}(d)}{t_d} \label{detun}
\end{eqnarray}
here $\delta_{\textbf{k}}$ is defined in Eqn.\ref{deltaM} in the section II as a $2\times2$ matrix in the spin space. But at $\textbf{k} = 0$, it turns out to be an identity matrix that can be treated as a number. $\tilde{\delta}$ essentially gives the energy difference between the electronic state in the thin film band closest(indexed by $n=N$) to the Dirac point and the Dirac point of the TI surface. When the energy difference is zero, the electrons at $k=0$ are in quantum-well resonance and the tunneling effect will be maximum. Moving away from $\tilde{\delta} = 0$ is equivalent to detuning away from resonance. We defined the detuning parameter at $\textbf{k}=0$ because we are mostly interested in studying the interacting dynamics of the electrons near the Dirac point. In general, one can define a detuning parameter for any general $\textbf{k}$. Here we use thin film thickness $d$ to tune the detuning parameter. 

From Eqns.\ref{Zfactor} and \ref{detun}, we could deduce the following simple relationship between  renormalization factors  and the dimensionless detuning parameter $\tilde{\delta}$ at zero momentum,
\begin{eqnarray}
    Z^t(\tilde{\delta}) = \frac{1}{2}\left( 1 + \frac{\tilde{\delta}}{\sqrt{1 + \tilde{\delta}^{2}}}\right) \,\, Z^b(\tilde{\delta}) = \frac{1}{2}\left( 1 - \frac{\tilde{\delta}}{\sqrt{1 + \tilde{\delta}^{2}}}\right)\nonumber\\ \label{Zdetun}
\end{eqnarray}

Fig.\ref{qwsproof} shows the results. In b) we plotted $Z^t$ and $Z^b$ as a function of the detuning parameter. a) part shows the band spectrum of the thin film and the TI surface at the three different limits of detuning.  When $\tilde{\delta} \ll 0$, $Z^b \approx 1$ and $Z^t \approx 0$. This implies that the bottom hybrid band is the thin film transverse band while the top band is the surface Dirac cone. On the other hand, when $\tilde{\delta} \gg 0$, the bottom band is the surface Dirac cone and the top band is the thin film transverse band. This is clearly understood once we look at the band dispersion shown in Fig.\ref{qwsproof}(a). In these two limits, the tunneling effects are perturbative. One can notice that the renormalization factor $Z^{b}$, which follows the surface band when $\tilde{\delta} \ll 0$. is nearly zero in this limit. Similar is the case with $Z^{t}$ when $\tilde{\delta} \gg 0$. This implies that the surface electrons do not experience a substantial attractive interaction when $|\tilde{\delta}| \gg 0$. 

But as $\tilde{\delta} \rightarrow 0$ from either side, things begin to change. We find that both the renormalization factors approach $1/2$ from either side. This clearly implies that the tunneling gets stronger and is non-perturbative. One can trace the surface Dirac cone by $Z^{b}$ when $\tilde{\delta} < 0$ and $Z^{t}$ when $\tilde{\delta} > 0$. We see that both the quantities rise up as $\tilde{\delta}$ approaches zero and reach a maximum equal to $1/2$ at $\tilde{\delta} = 0$. Recall that the interaction strength between the helical fermions is proportional to $Z^2$. Thus this spike at $\tilde{\delta} = 0$ is clear evidence of the surface fermions experiencing a maximum effective attractive interaction at $\tilde{\delta} = 0$.

On the other hand, the electrons that used to be in the thin film side when tunneling was zero now experience comparatively weaker attractive interaction. This is evident if we observe the evolution of $Z^{b}$ when $\tilde{\delta} < 0$ and $Z^{t}$ when $\tilde{\delta} > 0$. The two renormalization factors reach a minimum at $\tilde{\delta} = 0$ implying that the effective attractive interaction got weaker. 

In conclusion, by studying the evolution of the renormalization factors as a function of the detuning parameter, we showed that the effective attractive interaction acquired by the surface fermions near the Dirac point is the strongest when the thin film QW band is in quantum-well resonance with the surface Dirac cone. The fact that the $Z$-factors approach 1/2 at resonance suggests that there is no clear difference between the thin film fermions and the surface fermions at quantum-well resonance. This is clear evidence of our earlier proposition that the electronic states at the quantum-well resonance are hybridized. The eigenstates are a quantum superposition of the thin film and the surface states. They acquire a helical spin structure from the surface side and an effective attractive interaction between them from the thin film side. We shall be studying the superconductivity of these helical hybridized fermions within the BCS mean-field theory in the coming sections.

\section{\label{sec:level5}Effective mean-field Hamiltonian and the gap equation}
\subsection{Mean-field approximation}
Here we shall use the mean-field theory to decouple the four-fermion interaction Hamiltonian. Let $\triangle^{\text{hbd}}_{\alpha}(\textbf{k})$ be the order parameter on the helical hybrid band of index $m$($t$ or $b$) and helicity $\lambda$(=$+$ or $-$). Note that $\alpha = (m,\lambda)$. Similarly, define $\triangle^{\text{tf}}_{n}$ be the order parameter on the thin film band of index $n \neq N$. Now we apply mean-field approximation to the 4-fermion interaction Hamiltonian in Eqn.\ref{HI3},
\begin{widetext}
\begin{subequations}
\begin{eqnarray}
\triangle^{\text{hbd}}_{\textbf{k},\alpha} &=& \int \frac{d^{2}\textbf{p}}{(2\pi)^{2}}\left[\sum_{\beta = \{m',\lambda' \}} \lambda' e^{i\phi_{\textbf{p}}}J^{\alpha,\beta}_{\textbf{k},\textbf{p}}\left<a_{\textbf{p},\beta}a_{-\textbf{p},\beta}\right>  + \sum_{n\neq N}K^{n,\alpha}_{\textbf{k},\textbf{p}}\left< c^{T}_{\textbf{p},n}(is_y)c_{-\textbf{p},n}\right> \right] 
\end{eqnarray}
\begin{eqnarray}
    \triangle^{\text{tf}}_{\textbf{k},n} &=& \int \frac{d^{2}\textbf{p}}{(2\pi)^{2}} \left[\sum_{\alpha = \{m,\lambda \}}\lambda e^{i\phi_{\textbf{p}}}K^{n,\alpha}_{\textbf{k},\textbf{p}}\left<a_{\textbf{p},\alpha}a_{-\textbf{p},\alpha}\right> + \sum_{n'\neq N} V^{n,n'}_{\textbf{k},\textbf{p}}\left< c^{T}_{\textbf{p},n'}(is_y)c_{-\textbf{p},n'}\right> \right]
\end{eqnarray}    
\begin{eqnarray}
    \mathcal{H}_{\textbf{MF}} &=& \int \frac{d^{2}\textbf{k}}{(2\pi)^{2}} \left[\sum_{\alpha = \{m,\lambda \}} \lambda \triangle^{\text{hbd}}_{\textbf{k},\alpha} e^{-i\phi_{\textbf{k}}} a^{\dagger}_{\textbf{k},\alpha}a^{\dagger}_{-\textbf{k},\alpha} + \sum_{n \neq N}\triangle^{\text{tf}}_{\textbf{k},n} c^{\dagger}_{\textbf{k},n}(-is_y)c^{\dagger T}_{-\textbf{k},n} \right]
\end{eqnarray}
\end{subequations}    
\end{widetext}
Interestingly, the order parameters on the helical bands are of odd parity.
So we find that the helical fermions have an 'effective' p-wave pairing even though we started with a purely s-wave interaction. This is because the spin rotation symmetry(SRS) is broken by the induced spin-orbit coupling, while the time-reversal symmetry is preserved\cite{santos(2010),masatoshi(2011),nandkishore(2013)}. On the other hand, the pairing amplitude on the $n\neq N$ thin film transverse bands are of even parity. 

\subsection{The superconducting gap equation}

Using the mean-field theory, we derived the most general expression for the superconducting order parameter on the four helical hybrid bands and the remaining spin-degenerate thin film transverse bands. Note here that in our case, the fundamental origin of the attractive interaction is the electrons coupling to phonons. Since Debye frequency sets the UV cut-off for phonon modes, only electrons whose energy lies within the range $\left[\mu - \omega_{D}, \mu + \omega_{D}\right]$ can experience the attractive interaction. Here $\mu$ is the chemical potential. Here we focus on the limit $\omega_{D} \ll \mu$. This puts a strict constraint on the number of bands and the number of electrons participating in the pairing interaction. Only those bands that cross the Fermi level needed to be considered for pairing interaction. All those bands that lie above the Fermi level can be ignored. 
Before hybridization, the number of bands that cross the Fermi level can be calculated by taking the integer value of the expression, $ d/\pi \sqrt{2 m \mu/\hbar^2} + 1/2$. This integer will turn out to be the same as $N$, the index of the band that is hybridized with the surface Dirac cone. Hence before hybridization, we essentially have $2N$ Fermi surfaces because the thin film bands are spin-degenerate. 

\begin{figure}[b]
 \includegraphics[width=7.5cm, height=6cm]{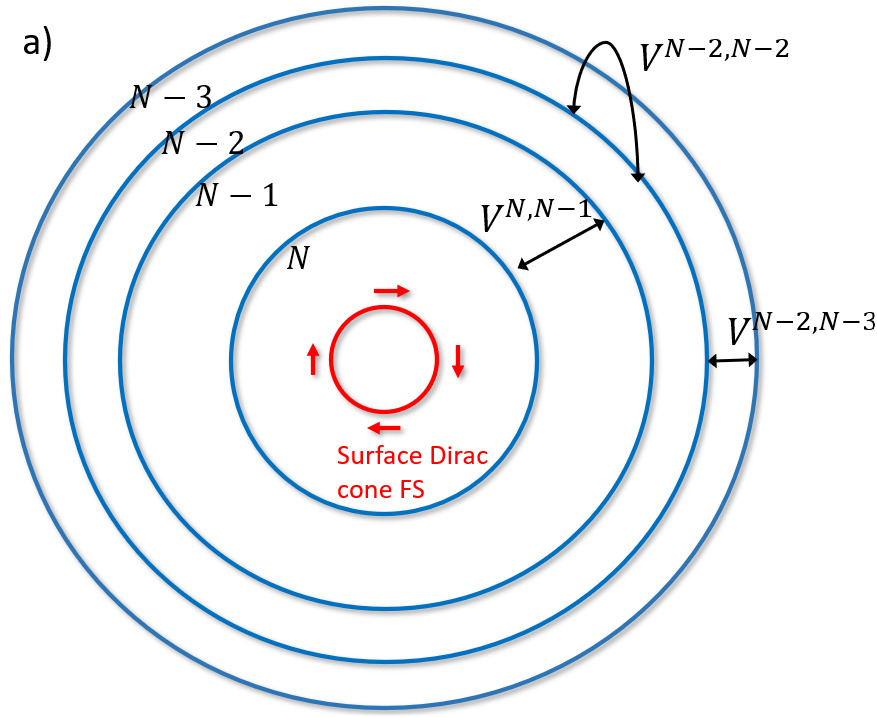} \\
 \includegraphics[width=7.5cm, height=6cm]{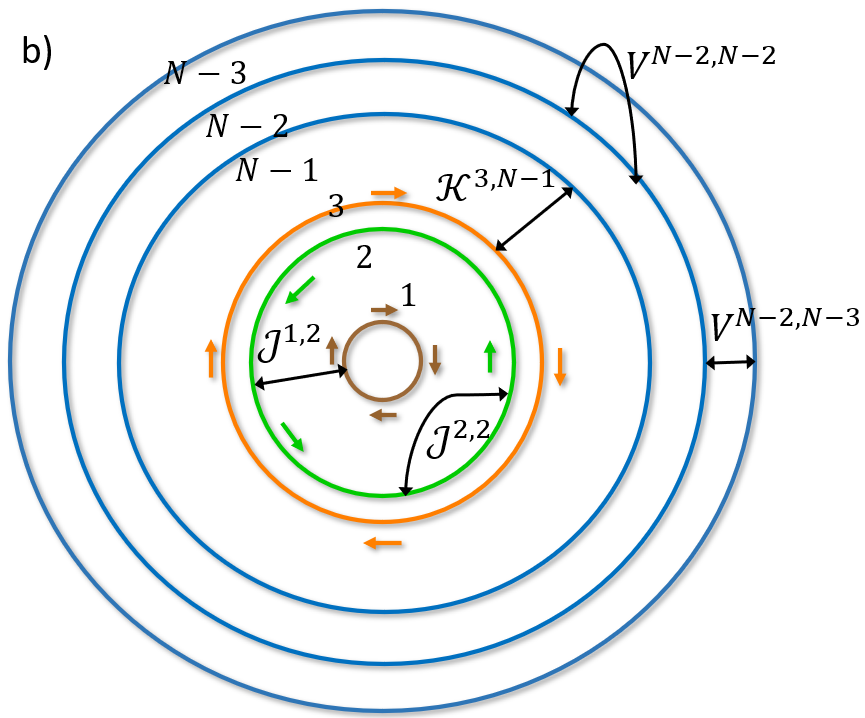}
 \caption{Schematic picture of the different Fermi surfaces in the thin film - TI hybrid before (a) and after (b) the tunneling is introduced. Also shown are the possible coupling matrix elements between the Fermi surfaces. In a), the blue circles represent the Fermi surfaces formed by the thin film QW bands, while the red circle corresponds to the Fermi surface formed by the surface Dirac cone.
 Each Fermi surface is twofold degenerate.  $V^{n,n'}$ gives the pairing matrix elements between the thin film Fermi surfaces. The electrons in the surface Dirac cone FS are non-interacting. After the tunneling is introduced in (b), $N$th band is hybridized with surface fermions. The tunneling effect on all the $n \neq N$ QW bands is ignored in our effective model given the condition in Eqn.\ref{condition} is satisfied. So the Fermi surfaces formed by all the $n < N$ QW bands remain unaffected by tunneling, hence spin-degenerate. $\mathcal{J}^{i,j}$ represents the coupling matrix element between the $i$th and $j$th hybrid FS. $\mathcal{K}^{i,n}$ represent the coupling matrix element between the $i$th hybrid FS and $n$th thin film Fermi surface. The definitions of the coupling matrix elements are given in Eqns.\ref{J},\ref{K},\ref{effcouple}.}
 \label{mmtncplng_scmtc}
 \end{figure}
 
 Now the chemical potential should be set within the bulk energy gap of the topological insulator. Once the thin film is deposited over the TI surface, the $N$th band is hybridized and we effectively have a 4-band model within the bulk gap. By fine-tuning the chemical potential further, it is possible that one can have the system with either three hybrid Fermi surfaces or just one Fermi surface (see fig.\ref{n=1limit}). In the latter case,  both the positive and negative helicity branches of the top band lie above the Fermi level and therefore do not participate in pairing. We shall derive the superconducting gap equation for these two cases separately here.

\subsubsection{3 hybrid Fermi surfaces + 2N - 2 thin film Fermi surfaces}
Now consider the case when the Fermi level is adjusted such that the hybrid has three Fermi surfaces within the thickness regime that we like to explore. We shall write down a gap equation for this specific case. The innermost Fermi surface(FS) was formed by the positive helicity branch of the $t$ (top) band while the next FS was formed by the negative helicity branch of the $t$ band. The outermost FS is formed by the positive branch of the $b$ (bottom) band. At this point, it is more convenient to express the superconducting gap and the coupling strength as functions of Fermi surface indices rather than the band indices.  In the weak-pairing limit ($\omega_D \ll \mu$), only electronic states very close to the Fermi surface take part in pairing. Thus, the electron renormalization factor that enters the pairing potential matrix can be re-expressed in terms of the Fermi momenta of the respective Fermi surfaces rather than the band indices. To support this, let us define three quantities $Z_1$, $Z_2$ and $Z_3$ for the three Fermi surfaces such that,
\begin{eqnarray}
    Z_{1} &=& Z^{(t,+)}_{\textbf{k}_{F1}},\,\,\,\,\,\,Z_{2} = Z^{(t,-)}_{\textbf{k}_{F2}}\nonumber\\ Z_{3} &=& Z^{(b,+)}_{\textbf{k}_{F3}} \label{Z}
\end{eqnarray}
where $1, 2$, and $3$ are the hybrid Fermi surface indices from smallest to largest in terms of size. Thus, $\textbf{k}_{F1}, \textbf{k}_{F2}$ and $\textbf{k}_{F3}$ are the Fermi momenta on these three hybrid Fermi surfaces.  Since the renormalization factor depends only on the magnitude of momentum, $Z_i$ is the same for all electrons in the Fermi surface indexed by $i$. The approximation we will do here is that we assume $Z_i$ factor is the same for all the electronic states lying within the energy window $[-\omega_D,\omega_D]$ measured from the chemical potential, given that the electronic states lie near the $i$th hybrid Fermi surface. This approximation allows us to re-express the interaction potential matrix in terms of the Fermi surface indices rather than the band indices. Let us define,
\begin{eqnarray}
 \mathcal{J}^{i,j}_{\textbf{k},\textbf{p}} = V^{N,N}_{\textbf{k},\textbf{p}} Z_i Z_j \label{J}
\end{eqnarray}
$\mathcal{J}^{ij}_{\textbf{k},\textbf{p}}$ is the interaction matrix element that gives the scattering strength of Cooper pair from the $i$th hybrid Fermi surface to the $j$th hybrid Fermi surface. 
One can also redefine $K^{n\alpha}_{\textbf{k},\textbf{p}}$ in terms of the Fermi surface indices. From Eqn.\ref{Z}, we have,
\begin{equation}    \mathcal{K}^{n,i}_{\textbf{k},\textbf{p}} = V^{n,N}_{\textbf{k},\textbf{p}} Z_{n}Z_{i}\label{K}
\end{equation}
where $\mathcal{K}^{n,i}_{\textbf{k},\textbf{p}}$ determines the scattering of Kramer's doublets from the $i$th hybrid Fermi surface to the $2n$th or $(2n - 1)$th ($n < N$) thin film Fermi surface. Here $2n$th and $(2n - 1)$th Fermi surfaces are formed by the helicity subbands of the $n$th spin-degenerate band. Due to this spin-degeneracy, the two helical Fermi surfaces overlap and hence the interaction parameters are the same for both. 

From the definition of $V^{N,N}_{\textbf{k},\textbf{p}}$ in Eqn.\ref{V}, we find that the matrix elements $\mathcal{J}^{i,j}_{\textbf{k},\textbf{p}}$ and $\mathcal{K}^{n,i}_{\textbf{k},\textbf{p}}$ are independent of momenta for electronic states lying within the Debye frequency measured from the Fermi level and zero otherwise. That is, we can write down the effective interaction potential in the following simple way,
\begin{eqnarray}
    V^{n,n'}_{\textbf{k},\textbf{p}} &=& V^{n,n'}\theta \left(\omega_D - \xi^{\text{tf}}_{\textbf{k},n}\right)\theta \left(\omega_D - \xi^{\text{tf}}_{\textbf{p},n'}\right)\nonumber\\ \mathcal{J}^{i,j}_{\textbf{k},\textbf{p}} &=& \mathcal{J}^{i,j} \theta \left(\omega_D - \xi^{\text{hbd}}_{\textbf{k},i}\right)\theta \left(\omega_D - \xi^{\text{hbd}}_{\textbf{p},j}\right)\nonumber\\
    \mathcal{K}^{n,i}_{\textbf{k},\textbf{p}} &=& \mathcal{K}^{n,i}\theta \left(\omega_D - \xi^{\text{tf}}_{\textbf{k},n}\right)\theta \left(\omega_D - \xi^{\text{hbd}}_{\textbf{p},i}\right)\label{effcouple}
\end{eqnarray}
where $\theta(x)$ is the Heavyside step function and the coupling matrix elements $V^{n,n'}$, $\mathcal{J}^{i,j}$ and $\mathcal{K}^{n,i}$ are independent of momenta. Also $\xi^{\text{tf(hbd)}}_{\textbf{k},n} = \epsilon^{\text{tf(hbd)}}_{\textbf{k},n} - \mu$ is just the energy of the thin film(hybrid) fermions measured from the chemical potential, involved in the interaction. 

With these definitions, it is straightforward to derive the superconducting gap equation. We shall also redefine the superconducting order parameters of the hybrid fermions also in terms of the Fermi surface indices as follows:
\begin{eqnarray}
   \triangle^{\text{hbd}}_{\textbf{k},1} &\approx& \triangle^{\text{hbd}}_{\textbf{k},t,+},\,\,\,\, \triangle^{\text{hbd}}_{\textbf{k},2} \approx \triangle^{\text{hbd}}_{\textbf{k},t,-}\nonumber \\ \triangle^{\text{hbd}}_{\textbf{k},3} &\approx& \triangle^{\text{hbd}}_{\textbf{k},b,+}
\end{eqnarray}
It has the form,
\begin{subequations}
\begin{eqnarray}
    \triangle^{\text{hbd}}_{\textbf{k},i} &-& \sum^{3}_{j=1}\int \frac{d^{2}\textbf{p}}{(2\pi)^{2}} \frac{\mathcal{J}^{i,j}_{\textbf{k},\textbf{p}}\triangle^{\text{hbd}}_{\textbf{p},j}}{2\sqrt{\left(\xi^{\text{hbd}}_{\textbf{p},j}\right)^2 + (\triangle^{\text{hbd}}_{\textbf{p},j})^2}}\nonumber\\ &=& \sum^{N-1}_{n=1}\int \frac{d^{2}\textbf{p}}{(2\pi)^{2}} \frac{\mathcal{K}^{n,i}_{\textbf{k},\textbf{p}}\triangle^{\text{tf}}_{\textbf{p},n}}{2\sqrt{(\xi^{\text{tf}}_{\textbf{p},n})^2 + (\triangle^{\text{tf}}_{\textbf{p},n})^2}}\label{gapeqtsc} \\
    \triangle^{\text{tf}}_{\textbf{k},n} &-& \sum^{N-1}_{n'=1} \int \frac{d^{2}\textbf{p}}{(2\pi)^{2}}\frac{V^{n,n'}_{\textbf{k},\textbf{p}}\triangle^{\text{tf}}_{\textbf{p},n'}}{2\sqrt{(\xi^{\text{tf}}_{\textbf{p},n'})^2 + (\triangle^{\text{tf}}_{\textbf{p},n'})^2}}\nonumber\\ &=& \sum^{3}_{i=1}\int \frac{d^{2}\textbf{p}}{(2\pi)^{2}}\frac{\mathcal{K}^{n,i}_{\textbf{k},\textbf{p}}\triangle^{\text{hbd}}_{\textbf{p},j}}{2\sqrt{\left(\xi^{\text{hbd}}_{\textbf{p},i}\right)^2 + (\triangle^{\text{hbd}}_{\textbf{p},i})^2}}\label{gapeqnormal}
\end{eqnarray}    
\begin{eqnarray}
    \xi^{\text{hbd}}_{\textbf{k},1} &=& \epsilon^{\text{hbd}}_{\textbf{k},t,+} - \mu, \,\,\,\, \xi^{\text{hbd}}_{\textbf{k},2} = \epsilon^{\text{hbd}}_{\textbf{k},t,-} - \mu\nonumber \\ \xi^{\text{hbd}}_{\textbf{k},3} &=& \epsilon^{\text{hbd}}_{\textbf{k},b,+} - \mu, \,\,\,\,  \xi^{\text{tf}}_{\textbf{k},n} = \epsilon^{\text{tf}}_{\textbf{k},n} - \mu \nonumber
\end{eqnarray}
\end{subequations}
With the weak-pairing approximation discussed above, the magnitude of the superconducting order parameters at all the Fermi surfaces turns out to be momentum-independent. The only possible momentum dependence on the gap magnitude could come from the restriction set by the Debye frequency. With this in mind, we shall define the parameters $\triangle^{\text{hbd}}_{i}$ and $\triangle^{\text{tf}}_{n}$ such that,
\begin{eqnarray}
    \triangle^{\text{hbd}}_{\textbf{k},i} &=& \triangle^{\text{hbd}}_{i} \,\theta \left(\omega_D - \xi^{\text{hbd}}_{\textbf{k},i}\right)\nonumber \\ \triangle^{\text{tf}}_{\textbf{k},n} &=& \triangle^{\text{tf}}_{n}\, \theta \left(\omega_D - \xi^{\text{tf}}_{\textbf{k},n}\right)
\end{eqnarray}
where $\theta(x)$ is the Heavyside step function. In all the future computations, we shall be representing the order parameters in dimensionless form as $\tilde{\triangle}^{\text{hbd}}_{i} = \triangle^{\text{hbd}}_{i}/\omega_D$ and $\tilde{\triangle}^{\text{tf}}_{n} = \triangle^{\text{tf}}_{n}/\omega_D$ where $\omega_D$ is the Debye frequency of the thin film metal.

A schematic picture of the coupling of Cooper pairs of electrons between different Fermi surfaces within the weak-coupling approximation before and after the tunneling is introduced is shown in fig.\ref{mmtncplng_scmtc}.

\subsubsection{1 hybrid Fermi surface + 2N - 2 thin film Fermi surfaces}

\begin{figure*}
 \includegraphics[width=14cm, height=8cm]{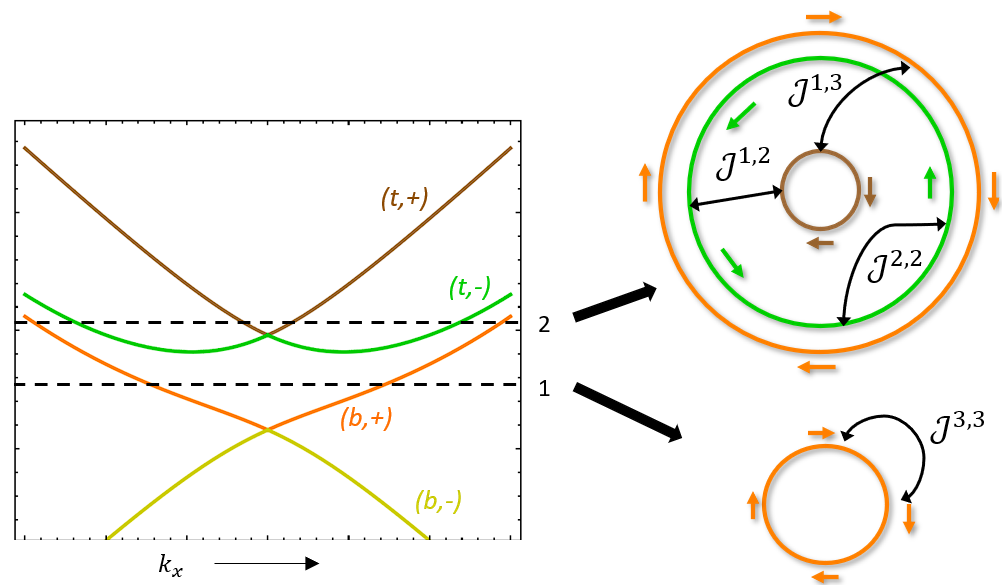} \\
 \caption{ Here we schematically show that by fine-tuning the chemical potential, it is possible to obtain cases with either 1) one Fermi surface or 2) three Fermi surfaces. The electrons on all these Fermi surfaces are helical, that is the spin-states are locked to their momentum direction. The arrowheads show the spin orientation of the Fermi electrons. $\mathcal{J}^{i,j}$ (where $i,j = 1,2,3$) represents the coupling matrix element between the Fermi surfaces indexed by $i$ and $j$. Its exact definition is given in Eqns.\ref{J}. For the single Fermi surface case, the Fermi level crosses the positive helicity branch of the 'bottom' band and the coupling matrix element is $\mathcal{J}^{3,3}$. Also shown here is the definition of hybridized band indices. Here $t(b)$ means the 'top'('bottom') band and '$\pm$' identifies the corresponding helicity.}
 \label{n=1limit}
\end{figure*}

Suppose that the Fermi level is fine-tuned to one hybrid Fermi surface within the bulk gap. That is, both the helicity branches of the top band are above the Fermi level (see fig.\ref{n=1limit}). Hence the top band does not contribute to the pairing at all. It is only the positive ( or the negative) helicity branch of the bottom band that crosses the Fermi level. One can observe that in the $N=1$ limit when there are no QW bands crossing the Fermi level, we effectively have a single band of helical fermions subject to attractive interaction. We shall study this limit more carefully in the next section. 

Since there is just one hybrid band crossing the Fermi level, the superconducting gap equation becomes far easier in this limit. Consider that it is the positive helicity branch of the $b$ band that crosses the Fermi level.  In this case, only the coupling constant $\mathcal{J}^{33}_{\textbf{k},\textbf{p}}$ survives. All the other elements vanish in this limit. For the interaction with thin film fermions, only $\mathcal{K}^{n,3}$ is needed to be taken into account. 
Hence in this limit,
the superconducting gap equation becomes,
\begin{subequations}
\begin{eqnarray}
    \triangle^{\text{hbd}}_{\textbf{k}} &-& \int \frac{d^{2}\textbf{p}}{(2\pi)^{2}} \frac{\mathcal{J}^{3,3}_{\textbf{k},\textbf{p}}\triangle^{\text{hbd}}_{\textbf{p}}}{2\sqrt{\xi^2_{\textbf{p},3} + (\triangle^{\text{hbd}}_{\textbf{p}})^2}}\nonumber\\ &=& \sum^{N-1}_{n=1}\int \frac{d^{2}\textbf{p}}{(2\pi)^{2}} \frac{\mathcal{K}^{n,3}_{\textbf{k},\textbf{p}}\triangle^{\text{tf}}_{\textbf{p},n}}{2\sqrt{(\xi^{\text{tf}}_{\textbf{p},n})^2 + (\triangle^{\text{tf}}_{\textbf{p},n})^2}}\label{gapeqtsc1FS} \\
    \triangle^{\text{tf}}_{\textbf{k},n} &-& \sum^{N-1}_{n'=1} \int \frac{d^{2}\textbf{p}}{(2\pi)^{2}}\frac{V^{n,n'}_{\textbf{k},\textbf{p}}\triangle^{\text{tf}}_{\textbf{p},n'}}{2\sqrt{(\xi^{\text{tf}}_{\textbf{p},n'})^2 + (\triangle^{\text{tf}}_{\textbf{p},n'})^2}}\nonumber\\ &=& \int \frac{d^{2}\textbf{p}}{(2\pi)^{2}}\frac{\mathcal{K}^{n,3}_{\textbf{k},\textbf{p}}\triangle^{\text{hbd}}_{\textbf{p}}}{2\sqrt{\xi^2_{\textbf{p},3} + (\triangle^{\text{hbd}}_{\textbf{p}})^2}}\label{gapeqnormal1FS}
\end{eqnarray}    
\end{subequations}
where $\triangle^{\text{hbd}}_{\textbf{k}} \approx \triangle^{\text{hbd}}_{\textbf{k},b,+}$. Here too we define $\triangle^{\text{hbd}}$ such that, 
\begin{equation}
\triangle^{\text{hbd}}_{\textbf{k}} = \triangle^{\text{hbd}} \theta(\omega_D - \xi^{\text{hbd}}_{\textbf{k}})    
\end{equation}

\section{\label{sec:level6}The N = 1 four band model}
 Here we shall present our work's simple yet most interesting result. Consider the case when the thin film transverse band of quantum number $N = 1$ is in resonance with the Dirac point of the topological insulator. Quantitatively from Eqns.\ref{HQW} and \ref{HTI2}, we find that the following condition should be satisfied: $\epsilon^{\text{tf}}_{0,n=1} = \epsilon^{\text{surf}}_{0,\pm}$. In other words, the detuning parameter $\tilde{\delta}(d) = 0$. If the material parameters of the topological insulator are fixed, then a practical way to achieve this condition is to tune the thin film thickness. So once the thickness is set and the thin film is deposited over the TI surface, the tunneling results in the hybridization of the electronic states near $\textbf{k} = 0$ resulting in the formation of four hybrid bands. Since we are in the $N=1$ limit, there are no trivial (or off-resonance) QW bands of index $n\neq N$ crossing the Fermi level. That is, only the hybridized fermions are present near the Fermi level. We know that the thin film favors an effective attractive interaction between electrons at zero temperature mediated by phonons. Therefore, we essentially have an effective model with helical hybridized fermions interacting via an effective attractive interaction between them. The full BCS interaction Hamiltonian in this $N=1$ limit attains the form,
 \begin{multline}
    \mathcal{H}_{I} = \mathcal{H}^{\text{hbd-hbd}}_{I}\\  = - \sum_{\alpha, \beta}\int \frac{d^{2}\textbf{k}}{(2\pi)^{2}}\frac{d^{2}\textbf{p}}{(2\pi)^{2}} e^{i\left(\phi_{\textbf{p}} - \phi_{\textbf{k}} \right)} \\ \lambda\lambda'J^{\alpha,\beta}_{\textbf{k},\textbf{p}}a^{\dagger}_{\textbf{k},\alpha}a^{\dagger}_{-\textbf{k},\alpha}a_{-\textbf{p},\beta}a_{\textbf{p},\beta}
 \end{multline}
 We have seen in the previous section that by fine-tuning the Fermi level, we essentially have phases with either three hybrid Fermi surfaces or just one hybrid Fermi surface as shown in Fig.\ref{n=1limit}. In this $N=1$ limit, these are the only Fermi surfaces present in the system. In the first part, we shall put forward the theoretical model in the two cases separately. In the last part, we shall tune various material parameters and look for possible enhancement of the superconducting gap. 

\subsection{Theoretical models}
\subsubsection{Single Fermi surface model}
\begin{figure}[b]
\includegraphics[width=6.cm, height= 4.3cm]{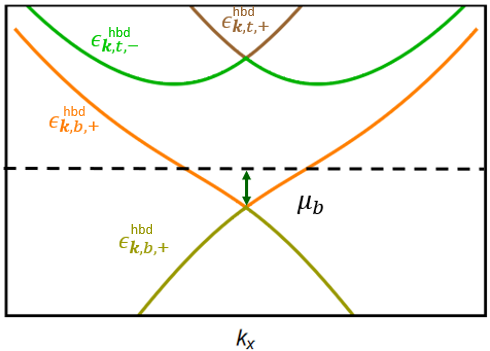}
\caption{\label{1fsspectra} Energy spectrum of the hybrid bands when the Dirac point of the TI surface is resonating with the $N=1$ thin film QW band. Here we show the case when the Fermi level(Black dashed lines) is tuned to a single Fermi surface and it crosses the positive helicity branch of the $b$ band. $\mu_b$ is the Fermi energy measured from the emergent Dirac point of the bottom band (band index - $(b,+)$). Its definition is given in Eqn.\ref{effchempot}.  }
\end{figure}

Here we consider the case when the Fermi level is tuned to one Fermi surface. This Fermi surface can be formed by either the positive or negative helicity branch of the bottom band. Since the interaction is mediated by the phonons, only the electronic states that lie within the energy window $\omega_{D}$ measured from the Fermi level experiences an attractive interaction. In this context, if the magnitude of the energy difference between the chemical potential and the emergent Dirac point of the bottom band is greater than the Debye frequency, then only the positive(negative) helicity states of the $b$ band experience attractive interaction. The negative(positive) branch is essentially non-interacting. Therefore, the projected Hamiltonian in the helicity basis resembles a single-band BCS problem for 'spinless fermions'. If the Fermi level crosses the positive helicity branch as shown in Fig.\ref{1fsspectra} the Hamiltonian attains the following simple form,
\begin{multline}
    \mathcal{H} =  \int \frac{d^{2}\textbf{k}}{(2\pi)^2} \biggl[a^{\dagger}_{\textbf{k},\alpha}\left[\epsilon^{\text{hbd}}_{\textbf{k},\alpha} - \mu\right]a_{\textbf{k},\alpha}\nonumber\\ - \int \frac{d^{2}\textbf{p}}{(2\pi)^2} \mathcal{J}^{3,3}_{\textbf{k},\textbf{p}} e^{i\left(\phi_{\textbf{p}} - \phi_{\textbf{k}} \right)} a^{\dagger}_{\textbf{k},\alpha}a^{\dagger}_{-\textbf{k},\alpha}a_{-\textbf{p},\alpha}a_{\textbf{p},\alpha}\biggr]
\end{multline}
where $\mathcal{J}^{3,3}_{\textbf{k},\textbf{p}}$ is defined in Eqn.\ref{J}. $\alpha = \{b,+\}$ is the band index.
 Following the procedure explained in Section V, mean-field Hamiltonian becomes,
 \begin{eqnarray}
     \mathcal{H}_{MF} &=&  \int \frac{d^{2}\textbf{k}}{(2\pi)^2}\left[ \triangle^{\text{hbd}}_{\textbf{k}} e^{-i\phi_{\textbf{k}}}a^{\dagger}_{\textbf{k},\alpha}a^{\dagger}_{-\textbf{k},\alpha} + h.c \right]\nonumber \\ \triangle^{\text{hbd}}_{\textbf{k}} &=& \int \frac{d^{2}\textbf{p}}{(2\pi)^2}\mathcal{J}^{3,3}_{\textbf{k},\textbf{p}}  e^{i\phi_{\textbf{p}}}\left<a_{\textbf{p},\alpha}a_{-\textbf{p},\alpha}\right>\nonumber \\ \mathcal{J}^{3,3}_{\textbf{k},\textbf{p}} &=& V^{N,N}_{\textbf{k},\textbf{p}} Z_3 Z_3 \label{n=1,1FS}
 \end{eqnarray}
 Here $N=1$ and $\mathcal{J}^{3,3}$ is the renormalized interaction potential between the helical fermions. Recall that $V^{N,N}$ is the thin film interaction potential matrix element between electrons in the $N$th band.  $Z_3$ is the renormalization factor of the electrons in the positive helicity branch of the bottom hybrid band at the Fermi momentum $k_F$. $Z_3$ essentially calculates the probability amplitude of a Kramer's pair of fermions to be in the thin film side of the interface. Since the hybrid fermions are a linear superposition of the thin film and the TI surface states, they acquire a helical spin texture from the TI surface side while also experiencing an effective attractive interaction mediated by the thin film phonons. The superconducting order is of odd parity as expected. 

 Here we shall present certain limits where simple analytical results for the superconducting gap can be derived. We will also show a limit where the effective pairing essentially goes back to singlet order. To identify these limits, let us define a parameter called $\mu_{b}$ with the following definition,
 \begin{equation}
     \mu_{b} = \mu - \epsilon^{\text{hbd}}_{0,b,+}\label{mudef}
 \end{equation}
 It is the difference in energy between the Fermi level and the emergent Dirac point of the bottom band. $\mu_{b} = 0$ implies the Fermi level is aligned with the Dirac point and the Fermi surface reduces to just a Fermi point. So one can call this term an 'effective' chemical potential of the bottom band. Let us represent $\mu_{b}$ in dimensionless form by dividing it with the tunneling strength $t_d$ defined in Eqn.\ref{Ht}. That is, 
 \begin{eqnarray}
     \tilde{\mu}_{b} = \frac{\mu_{b}}{t_d} \label{effchempot}
 \end{eqnarray}
 Here the thickness $d$ is fixed. When $\tilde{\mu}_b \ll 1$, we find that the energy dispersion of the states that cross the Fermi level is essentially a linear function of $\textbf{k}$. That is, the energy of Fermi electrons can be approximated as,

 \begin{eqnarray}
\epsilon^{\text{hbd}}_{\textbf{k},b,+} - \mu \approx +A_b |\textbf{k}| - \mu_b     
 \end{eqnarray}
 $A_b = A_0/2$ is the effective spin-orbit coupling on the helical fermions in the $b$ band near the Dirac point. When the Debye frequency $\omega_D < \mu_b$, only the positive helicity branch is interacting. In this limit, one can solve Eqn.\ref{n=1,1FS} analytically to arrive at a simple expression for the magnitude of the p-wave pairing gap,
 \begin{eqnarray}
     \triangle^{\text{hbd}} = 2 \omega_D\text{Exp}\left[-\frac{4\pi A^2_b}{\mu_b \mathcal{J}^{3,3}} \right]\label{lineardirac}
 \end{eqnarray}
  Note here that if $\mu_{b}<\omega_{D}$, then both the negative and the positive helicity branches of the $b$ band fall within the energy window $\left[\mu_b - \omega_D, \mu_b + \omega_D \right]$. This implies that the electronic states of both helicities that fall within this window will be interacting. The effective theory described in Eqn.\ref{n=1,1FS} does not explain the full physics in this limit.

 A rather interesting limit is when the chemical potential $\mu_b = 0$. In this limit, hybrid electronic states of both the helicity branches experience attractive interaction on an equal footing. Therefore, the triplet component of the order parameter cancels out. That is, we essentially have a purely singlet-pairing superconducting phase of helical Dirac fermions. In the limit when $\omega_D \ll t_d$, the effective low-energy interacting Hamiltonian in this limit has the form:
 \begin{eqnarray}
     \mathcal{H} &=&  \int \frac{d^{2}\textbf{k}}{(2\pi)^2} \,\biggl[A_b\,d^{\dagger}_{\textbf{k},b}\left[ \textbf{s}\times \textbf{k}.\hat{z} \right]d_{\textbf{k},b}\nonumber \\ &-&  \int \frac{d^{2}\textbf{p}}{(2\pi)^2}\,\mathcal{V}_{\textbf{k},\textbf{p}}\, d^{\dagger}_{\textbf{p},b} s_y d^{\dagger T}_{-\textbf{p},b} d^{T}_{-\textbf{k},b}s_y d_{\textbf{k}, b} \biggr]\\ 
     \mathcal{V}_{\textbf{k},\textbf{p}} &\approx& \frac{V^{1,1}_{\textbf{k},\textbf{p}}}{4}  \nonumber
 \end{eqnarray}
  where $V^{1,1}_{\textbf{k},\textbf{p}}$ is the thin film phonon-mediated interaction potential between the electronic states in the transverse bands indexed by $N = 1$. Its definition is given in Eqn.\ref{V}. The factor of $4$ is because in the limit when $\omega_D \ll t_d$,  the renormalization factor is diagonal in the spin basis with both the diagonal elements equal to $1/2$. In other words, the electrons involved in the interaction are in quantum-well resonance.  
  $d_{\textbf{k},b} = \begin{bmatrix} d_{\textbf{k},b,\uparrow} & d_{\textbf{k},b,\downarrow} \end{bmatrix} $ is the 2-component spinor representing the annihilation operator for emergent Dirac fermions of the $b$ band in the spin basis. This effective theory has extra emergent symmetries in contrast to the finite chemical potential case. One can see that it has both the particle-hole symmetry and the Lorentz symmetry. Since there are no Fermi electrons in this limit to induce Cooper instability, the coupling constant must be greater than a critical value for the superconducting phase transition to happen\cite{nandkishore(2013)}. The critical value of the interaction strength is given by,
  \begin{equation}
      \mathcal{V}_c = \frac{4\pi A^2_b}{\omega_D}
  \end{equation}
  If the interaction strength is tuned to the quantum critical point, the effective theory possesses emergent surface supersymmetry(SUSY). So what we have here is essentially a very practical platform to study the dynamics of the emergent supersymmetric quantum matter.

\subsubsection{Three Fermi surface model}
\begin{figure}[b]
\includegraphics[width=6.cm, height= 4.96cm]{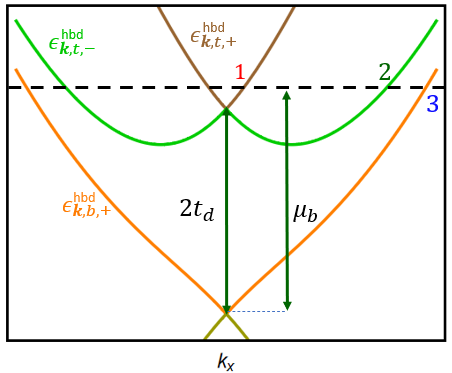}
\caption{\label{3fsspectra} Similar setup as in Fig.\ref{1fsspectra} but here we tuned the Fermi level to three Fermi surfaces. $t_d$ is the tunneling strength (defined in Eqn.\ref{effchempot}). The numbers here represent the Fermi surface indices.}
\end{figure}
Now consider the case when the Fermi level is adjusted in such a way that we effectively have three Fermi surfaces. A schematic picture of such a possibility is shown in Fig.\ref{n=1limit}. To realize a three Fermi surface model, the effective chemical potential of the $b$ band defined as $\mu_b$ in Eqn.\ref{mudef} has to be greater than $2t_d$. 
 In this limit, the Fermi surface closest to the Dirac point is formed by either positive or negative helicity branches of the $t$ band depending on the fine-tuning of the chemical potential. This is indexed by $1$. The second and third Fermi surfaces are formed by the negative helicity branch of the $t$ band (band index - $(t,-)$)  
 and the positive helicity branch of the $b$ band (band index - $(b,+)$) respectively. They are indexed as $2$ and $3$ respectively. Since the attractive interaction is mediated by phonons, only the electronic states lying within the energy window $\pm \omega_D$ measured from the chemical potential actually experience an attractive interaction. Since we are working in the limit where $\omega_{D} \ll \mu$, the absolute value of the chemical potential, essentially only the electrons in and around the Fermi level take part in the interaction. Also note that, since we are in the $N=1$ limit, only the helical hybrid fermions are present in the system. The mean field Hamiltonian then takes the form,
\begin{eqnarray}
    \mathcal{H}_{\text{MF}} &=&  \int \frac{d^{2}\textbf{k}}{(2\pi)^2}\biggl[ \triangle^{\text{hbd}}_{\textbf{k},1} e^{-i\phi_{\textbf{k}}}a^{\dagger}_{\textbf{k},t,+}a^{\dagger}_{-\textbf{k},t,+}  \nonumber \\ &-&  \triangle^{\text{hbd}}_{\textbf{k},2} e^{-i\phi_{\textbf{k}}}a^{\dagger}_{\textbf{k},t,-}a^{\dagger}_{-\textbf{k},t,-}\nonumber \\ &+&   \triangle^{\text{hbd}}_{\textbf{k},3} e^{-i\phi_{\textbf{k}}}a^{\dagger}_{\textbf{k},b,+}a^{\dagger}_{-\textbf{k},b,+} + h.c \biggr]\nonumber\\     
\end{eqnarray}
Here we assumed that the Fermi level crosses the positive helicity branch of the top band to form the Fermi surface that is closest to the Dirac point as shown in Fig.\ref{3fsn=1}(a). Also, the energy difference between the Dirac point of the $t$ band and the Fermi level must be higher than the Debye frequency for the above Hamiltonian to effectively describe the pairing physics. Otherwise, the electrons in the negative helicity branch of the $t$ band near $k=0$ will also be interacting. This is not taken into account in the effective Hamiltonian defined here. 

As long as the three hybrid Fermi surfaces do not overlap in the momentum space, the superconducting order on each of them is of p-wave symmetry. Notice that since the Fermi surface indexed by $2$ is formed by the negative helicity branch of the $t$ band, the sign of the order parameter is negative. That is, it differs from the order parameter on the positive helicity branch by a phase of $\pi$. If this Fermi surface happens to overlap with a positive helicity branch of the $b$ band, which could happen in case the tunneling is zero or negligible, then one can find that the triplet component of the order parameter cancels out. In that case, we are left with an even-parity spin-singlet pairing phase. 

The superconducting gap equation satisfied by $\triangle^{\text{hbd}}_{i}$'s is similar to what is given in Eqn.\ref{gapeqtsc}. But since there are no thin film FSs, the RHS of Eqn.\ref{gapeqtsc} vanishes. So we finally obtain a simple form for the gap equation which we shall write down below for clarity,
\begin{eqnarray}
        \triangle^{\text{hbd}}_{\textbf{k},i} &-& \sum^{3}_{j=1}\int \frac{d^{2}\textbf{p}}{(2\pi)^{2}} \frac{\mathcal{J}^{ij}_{\textbf{k},\textbf{p}}\triangle^{\text{hbd}}_{\textbf{p},j}}{2\sqrt{\xi^2_{\textbf{p},j} + (\triangle^{\text{hbd}}_{\textbf{p},j})^2}}\nonumber \\ &=& 0 \label{gapeq3fsn=1}
\end{eqnarray}
where $i=1,2,3$. The matrix elements of $\hat{J}$ are given in Eqn.\ref{J}. It describes the scattering strength of Kramer's doublets from the Fermi surface indexed by $i$ to $j$. 

So we find here that we have to effectively solve a set of 3 non-linear coupled integral equations to find the superconducting order parameters in each Fermi surface. A simple analytical solution as was done in the single Fermi surface case is difficult to realize here. 
\subsection{Numerical results: Solving the gap equation}

The objective of this part of the section is to study the evolution of the superconducting order in the $N=1$ limit as a function of various tuning parameters. Basically, our goal is to look for various ways to enhance the superconductivity. The role of the thin film in this hybrid system is to induce an effective attractive interaction between the helical surface fermions. Therefore, a straightforward way to enhance the pairing interaction between the helical hybrid fermions will be to tune the electron-phonon coupling strength of the thin film metal. In the case of a topological insulator, it is the spin-orbit interaction that decides the Fermi velocity of the surface Dirac fermions. So understanding the evolution of the superconducting order as a function of the spin-orbit coupling strength is important.

Here we begin by  
emphasizing again the role played by quantum-well resonance in realizing a ground state with attractively interacting helical fermions and in enhancing the superconducting order. This is a continuation of the physics discussed in section IV. There we discussed how the effective attractive interaction attained by the surface fermions through tunneling reaches its maximum when the two systems are in quantum-well resonance. We used the evolution of the $Z$-factors of the two hybrid bands as a function of the detuning parameter to prove this point. Having derived the pairing gap equation, we can finally study how the pairing gap on the Fermi surfaces evolves as a function of the detuning parameter. This will give a rather concrete idea of why we must tune the thin film thickness to quantum-well resonance for a given $N$ to study the interacting physics of surface fermions.

In short, we essentially write down the p-wave superconducting gap on the hybrid bands as a function of the three tuning parameters,
\begin{eqnarray}
    \triangle^{\text{hbd}}_{i} \equiv \triangle^{\text{hbd}}_{i}(\tilde{\delta}(d), \tilde{\lambda}^{\text{bulk}}, \tilde{v})
\end{eqnarray}
Here $\tilde{\lambda}^{\text{bulk}}$ is the dimensionless form of the phonon-mediated interaction strength of the 3D bulk counterpart of the metal thin film. In terms of the electron-phonon coupling strength $G_{fp}$ defined in Eqn.\ref{eph}, \begin{eqnarray}
    \tilde{\lambda}^{\text{bulk}} = \frac{m k^{\text{bulk}}_{F}}{2\pi^2\hbar^2} G^2_{\text{fp}}\label{bulklambda}
\end{eqnarray} 
$k^{\text{bulk}}_{F}$ is the bulk Fermi momentum of the metal for a given chemical potential. In the calculations here, we shall only tune the electron-phonon coupling strength of the metal while keeping all other parameters constant. The dimensionless detuning parameter is defined in Eqn.\ref{detun}. $\tilde{v}$ here is the dimensionless form to represent the Fermi velocity of the surface fermions. For the class of topological insulators that we consider, it is proportional to the SOC strength of the TI. It has the following definition,
\begin{equation}
    \tilde{v} = \frac{A_0}{\hbar c}\label{soc}
\end{equation}
where $A_0$ is the SOC strength of the topological insulator and $c$ is the speed of light. Tuning down $\tilde{v}$ is essentially equivalent to moving towards the flat band limit of the TI surface.
Now we shall study the evolution of the superconducting order as a function of these dimensionless tuning parameters. For numerical purposes, we shall be using material parameters corresponding to Pb(lead) for the thin film except in the section where we tune the interaction strength.
\begin{figure*}
    \includegraphics[width=11cm, height=7.cm]{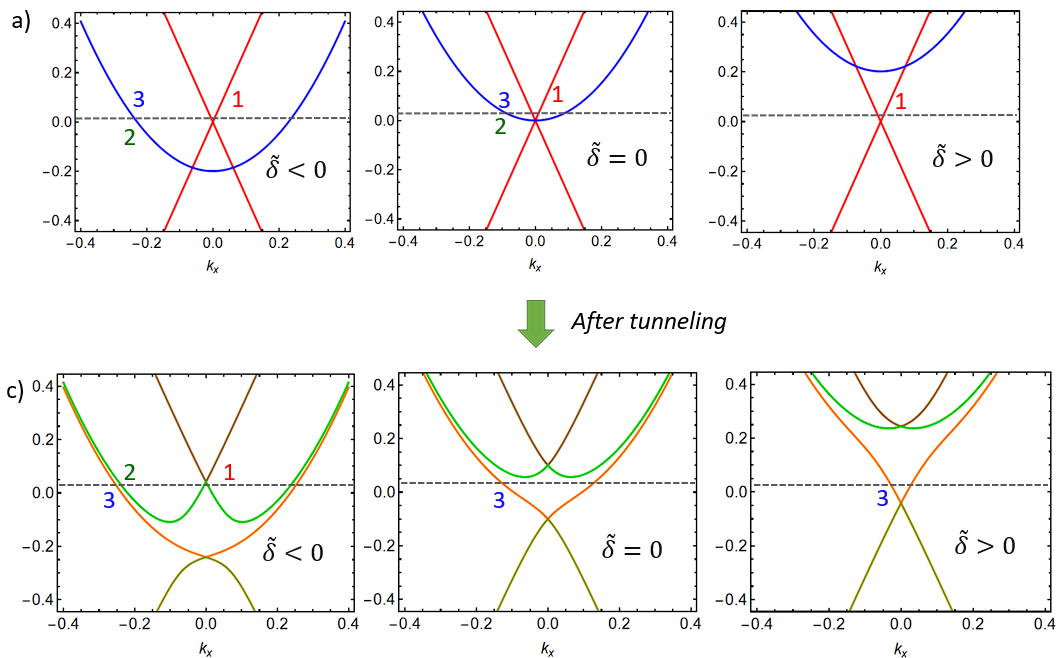} \includegraphics[width=5cm, height=7.7cm]{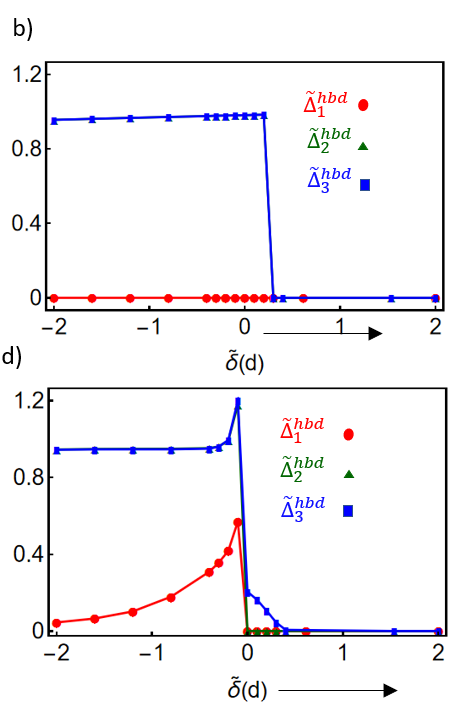} \\
 \caption{Resonance effect: Plots showing the emergence of topological superconductivity on the thin film-TI hybrid due to effective attractively interacting surface fermions when the detuning parameter $\tilde{\delta}(d)$ approaches zero. (a) shows the energy spectra of the TI surface Dirac cone and the thin film transverse band at three regimes of the detuning parameter before the tunneling is turned on. (c) shows the energy spectra of the hybrid bands at the same three values of the detuning parameter after the tunneling is turned on. The grey horizontal line denotes the Fermi level. The numbers denote the Fermi surface indices as defined in the main text. (b) shows the evolution of the superconducting order parameters $\triangle^{\text{hbd}}_{i}$'s on the three Fermi surfaces indexed by $i=1,2,3$ as a function of the detuning parameter before tunneling is turned on. (d) shows the evolution of the same set of order parameters after tunneling is turned on.}
 \label{qwspairing}
\end{figure*}
\subsubsection{Resonance effect}
Here we shall study the evolution of the p-wave pairing gaps as a function of the dimensionless detuning parameter at $\textbf{k} = 0$ defined in Eqn.\ref{detun}. The detuning parameter is varied by tuning the thin film thickness. We shall solve the gap equation both before and after the tunneling is turned on. The Fermi level is set at $0.05 \text{eV}$ above the Dirac point of the topological insulator. Essentially, we set the Fermi level close to the Dirac point because we are tuning the detuning parameter defined at $k=0$. If the Fermi level is much above or below the Dirac point, then the detuning parameter should be defined at the Fermi momentum instead of at $k=0$.

Fig.\ref{detun} shows the results. Here the detuning parameter is varied from $-2$ to $2$. We have studied the evolution of the pairing gaps $\triangle^{\text{hbd}}_{1}$(Red), $\triangle^{\text{hbd}}_{2}$(Green) and $\triangle^{\text{hbd}}_{3}$(Blue)  on the three Fermi surfaces(if present) before and after the tunneling is switched on. Before the tunneling is turned on, the innermost Fermi surface is formed by the surface Dirac cone. The second and third Fermi surfaces are formed entirely by the two helicity branches of the thin film band and hence they overlap. Essentially in this limit, the TI surface is non-interacting, which means we are studying just the thin film superconductivity. The purpose is just to set a benchmark for the study of the superconducting order once the tunneling is turned on.  Therefore, $\triangle^{\text{hbd}}_{1}$ is always zero. And we have $\triangle^{\text{hbd}}_{2} = \triangle^{\text{hbd}}_{3}$. The triplet component of the order parameter cancels out and we have the trivial s-wave superconducting order as expected. When the detuning parameter is increased, the thin film band starts moving up. This is because, in our convention, increasing the detuning parameter is equivalent to reducing the thin film thickness. At a particular thickness, the bottom of the band crosses the Fermi level. Beyond this point, there are no interacting Fermi electrons. Hence superconductivity vanishes as the detuning parameter is increased further. 

Now when the tunneling is turned on, the surface band and the thin film band get hybridized. From the figure, we understand that the pairing physics is not very different from the zero-tunneling result when $|\tilde{\delta}| \gg 0$. But as we fine-tune $\tilde{\delta}$ to zero, we start seeing the effects of electronic hybridization. The electrons in the innermost Fermi surface, which essentially is the surface Dirac cone start interacting and a superconducting gap opens up. The magnitude of the gap increases as we fine-tune to $\tilde{\delta} = 0$ from the left side. One can identify that $\triangle^{\text{hbd}}_{1}$(the red points in the plot) is the effective pairing gap on the Dirac cone. Note that the contribution to the pairing gap also comes from the scattering of Cooper pairs to the other two Fermi surfaces as well. 

When the detuning parameter is increased further, the bottom of the $t$ hybrid band crosses the  Fermi level. This means, there is essentially a crossover from the three Fermi surface to the single Fermi surface limit. Both the 1st and the 2nd Fermi surfaces vanish beyond this limit. When the tunneling was zero, there was no superconductivity in this limit because the surface was essentially non-interacting. But here we see that a superconducting gap exists on the Fermi surface formed by the Dirac cone (the blue-colored points on the plot). This is clear evidence of the effective attractive interaction between the surface Dirac fermions. Also, we see that the magnitude of the gap decreases as the detuning parameter is tuned away from zero.
This clearly proves that the quantum-well resonance is the ideal point to study the attractive interacting physics of surface Dirac fermions.

\begin{figure}[b]
\includegraphics[width=7.cm, height=4.5cm]{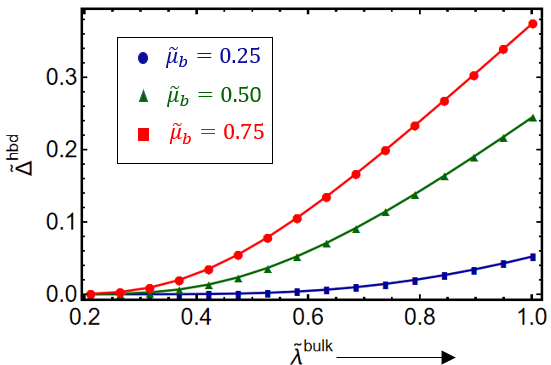}
\caption{\label{interac1fs} Single hybrid Fermi surface limit (see fig.\ref{1fsspectra} for the energy spectrum): Plot showing the evolution of the superconducting order parameter of the resonating hybrid electrons as a function of the bulk coupling constant of the thin film $\tilde{\lambda}^{\text{bulk}}$(defined in Eqn.\ref{bulklambda}). The evolution is studied at three different values of the effective chemical potential expressed in a dimensionless form as $\tilde{\mu}_b$(defined in Eqn.\ref{effchempot}). The SOC strength is fixed at $A_0 = 1.5 \text{eV}$\AA and the tunneling strength $t_d = 0.2 \text{eV}$.}
\end{figure}

\subsubsection{Dependence on the interaction strength}

In part 1, we understood the importance of quantum-well resonance to realize a phase with attractively interacting helical surface fermions. So from here onwards, we fine-tune the thickness to quantum-well resonance at the Dirac point. In this limit, the electronic states close to the Dirac point on both sides of the interface are strongly hybridized. There is no clear difference between the thin film and the TI surface fermions. These resonating hybrid fermions acquire the emergent spin-orbit coupling from the thin film side and an effective attractive interaction from the thin film side. We effectively have helical fermions with an effective attractive interaction between them.

Here we tune the electron-phonon coupling strength $G_{\text{fp}}$ of the thin film metal and study the evolution of the pairing gap on the hybrid Fermi surfaces. To represent the tuning parameter in a dimensionless form, we defined the bulk coupling constant of the metal $\tilde{\lambda}^{\text{bulk}}$ in Eqn.\ref{bulklambda}. We keep all other material parameters including Debye frequency, effective electron mass, etc. constant. Here we used the material parameters of the Pb metal for numerical calculations. The cases of single and three Fermi surfaces were considered separately. The effective chemical potential was fine-tuned further for each of the two cases to understand its significance.

 \paragraph{Single Fermi surface}Fig.\ref{interac1fs} shows the results in the case when the chemical potential is tuned to a single Fermi surface. Here we plotted the magnitude of the p-wave superconducting gap represented in a dimensionless form(with respect to the Debye frequency) at three different chemical potential values, $\tilde{\mu}_b = 0.25, 0.50, 0.75$. Here chemical potential is expressed in a dimensionless form as $\tilde{\mu}_b = \mu_b/t_d$ where the tunneling strength $t_d$ is fixed at $t_d = 0.2 \text{eV}$. The chemical potential is set very close to the Dirac point because the Fermi electrons then will be at quantum well resonance. In addition, the electron band will be linear, resembling a surface Dirac cone. The corresponding energy spectrum is shown in fig.\ref{1fsspectra} We set the spin-orbit coupling strength at $A_0 = 1.5 \text{eV}$\AA. To arrive at this result, we numerically solved Eqn.\ref{n=1,1FS} self-consistently at different values of the coupling strength.

\begin{figure*}
 \includegraphics[width=16cm, height=5cm]{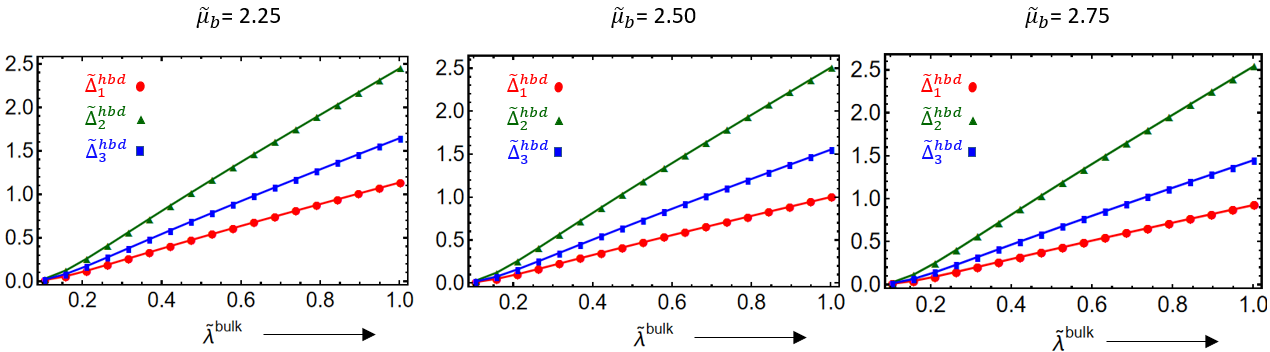} \\
 \caption{Three hybrid Fermi surfaces (see fig.\ref{3fsspectra} for the energy specctrum): Similar setting as in fig.\ref{interac1fs} but here the effective chemical potential $\tilde{\mu}_b > 2$ for all the three cases. Thus we have three hybrid Fermi surfaces. The evolution of the SC order parameters on these three Fermi surfaces has been studied as a function of the bulk interaction strength of the thin film.}
 \label{interac3fs}
\end{figure*}

As expected, we find an exponential enhancement of the superconducting gap as the coupling constant $\tilde{\lambda}^{\text{bulk}}$ is increased. Increasing chemical potential also enhances the superconducting gap. The results can be explained in the following way: Since the chemical potential is set close to the Dirac point($\tilde{\mu}_{b} < 1$), the band is nearly linear when it crosses the Fermi level. Hence the approximate analytical expression for the pairing gap magnitude derived in Eqn.\ref{lineardirac} works well in these cases. There we found that $\triangle^{\text{hbd}} \propto e^{-1/\mu_b \mathcal{J}^{3,3}}$. Here $\mathcal{J}^{3,3}$ is proportional to the electron-phonon coupling constant. Thus both the chemical potential and the interaction strength have a similar enhancement effect on the superconducting gap magnitude. This is in contrast to a 2D quadratic electronic dispersion. There the density of states is independent of the chemical potential. Note here that, if $\mu_b \ge t_d$, then the band is no longer linear. In this case, the analytical result derived in Eqn.\ref{lineardirac} is no longer a good approximation. In addition, since the Fermi electrons lie away from the quantum-well resonance, the tunneling effects will be perturbative.

\paragraph{Three Fermi surfaces} Fig.\ref{interac3fs} shows the results when the chemical potential is tuned to three Fermi surfaces. As discussed before, the effective chemical potential, $\mu_b$ must be of the order of $2t_d$ or greater than that to realize a three Fermi surface model. The corresponding energy spectrum is given in Eqn.\ref{3fsspectra} Here we studied the evolution of the p-wave pairing gaps on the three Fermi surfaces as a function of the electron-phonon coupling strength of the thin film metal at three different values of the chemical potential. Here $\triangle^{\text{hbd}}_{i}$(i=1,2,3) is the SC gap magnitude on the $i$th Fermi surface. Here $1$ is the closest and $3$ is the farthest from the Dirac point. They are represented in a dimensionless form by dividing them with the Debye frequency of the thin film metal. We used the dimensionless parameter $\tilde{\mu}_b$ to represent the chemical potential. The tunneling strength and the spin-orbit coupling strength of the TI surface are all fixed with the same numerical values as in the single Fermi surface case. We numerically solved the coupled set of superconducting gap equations given in Eqn.\ref{gapeq3fsn=1} to arrive at these results. 

We see a much-anticipated enhancement in the superconducting gap magnitude as the interaction strength is increased. We also  notice that the magnitude of the superconducting gap is substantially larger compared to the single Fermi surface case for a given strength of interaction. This is because there is a larger number of Fermi electrons involved in the interaction for the three Fermi surface cases, leading to an enhancement in the superconducting order.

\subsubsection{Dependence on the spin-orbit coupling strength}

Here we shall study the evolution of the superconducting order on the helical hybrid bands as a function of the spin-orbit coupling strength of the TI surface. As we did in the previous part, the Dirac point of the TI surface is fixed at quantum-well resonance with the $N=1$ transverse band of the thin film. The bulk interaction strength
is fixed at $\tilde{\lambda}^{\text{bulk}} = 0.39$. As before, the tunneling strength is fixed at $t_d = 0.2 \text{eV}$. The spin-orbit coupling strength is expressed in a dimensionless form given by $\tilde{v} = A_0/\hbar c$. The logic here is that for a given SOC strength $A_0$, the Dirac velocity of the surface fermions is given by $v = A_0/\hbar$. So tuning the SOC strength is equivalent to tuning the Dirac velocity of the surface fermions.

We study the SOC dependence for the two cases separately: when the Fermi level is set to a single Fermi surface and when the Fermi level is set to three Fermi surfaces. Even though we expect a monotonic increase in the superconducting gap as the SOC strength is decreased due to the obvious increase in the density of states, we shall find here that it is not the case. The change in the hybrid band structure has huge consequences on the renormalization factors $Z_i$  which substantially affects the pairing interaction. 
\begin{figure*}
 \includegraphics[width=12cm, height=9cm]{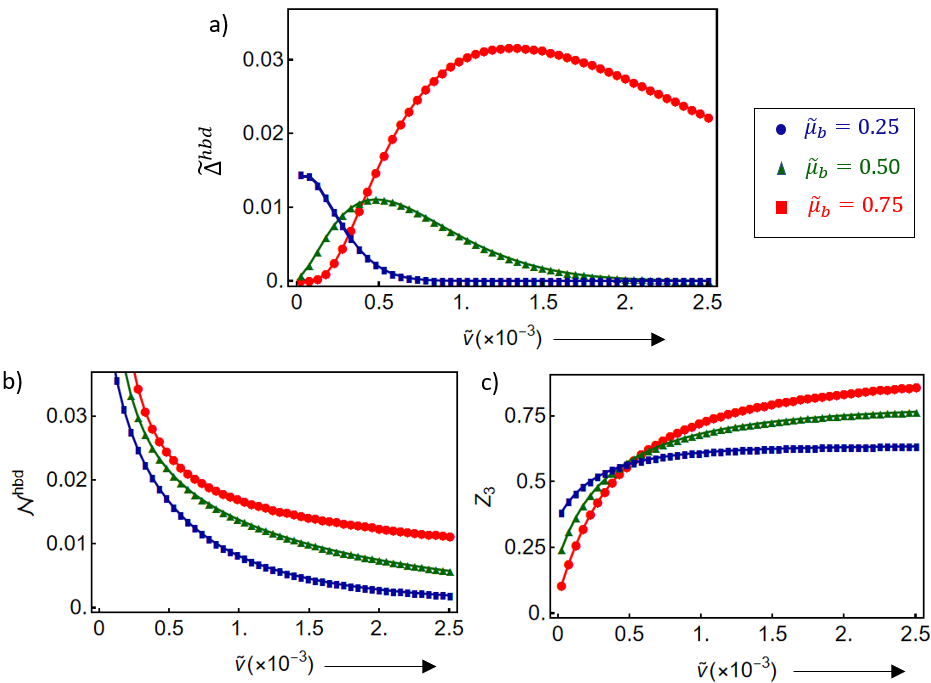} \\
 \caption{Single Fermi surface model (See fig.\ref{1fsspectra}): Plot (a) shows the evolution of the superconducting order parameter as a function of $\tilde{v}$. Here $\tilde{v} = \frac{A_0}{\hbar c}$ is used to represent the spin-orbit coupling strength of the TI in a dimensionless way. It also indicates the Dirac velocity of the surface fermions. We study the pairing at three different values of the effective chemical potential $\tilde{\mu}$(Definition given in Eqn.\ref{effchempot}). (b) shows the evolution of the density of states at the Fermi level as $\tilde{v}$ is tuned. (c) shows the evolution of the $Z$-factor of the hybrid Fermi electronic states involved in pairing. It represents the probability amplitude of the Fermi electrons to be in the thin film side of the interface. The expression is given in Eqn.\ref{Z}. The tunneling strength is fixed at $t_d = 0.2 \text{eV}$.}
 \label{1fsresult}
\end{figure*}

\paragraph{Single Fermi surface} Here we shall study the evolution of the pairing gap as a function of the spin-orbit coupling parametrized by $\tilde{v}$ at different values of $\tilde{\mu}_{b}$. Since the magnitude of the superconducting gap in our case is mostly decided by the density of states at the Fermi level and the renormalization factor $Z_3$, we have plotted both of them 
 as a function of $\tilde{v}$. This helps us better understand the behavior of $\tilde{\triangle}^{\text{hbd}}$ as $\tilde{v}$ is tuned. The density of states at the Fermi level has the following definition,
\begin{eqnarray}
    \mathcal{N}^{\text{hbd}} = \int \frac{d^{2}\textbf{k}}{(2\pi)^2} \delta(\epsilon^{\text{hbd}}_{\textbf{k},b,+} - \mu)
\end{eqnarray}By tuning $\tilde{v}$, we shall expect the density of states at the Fermi level to increase thus enhancing the superconductivity. But here we shall find that it is not always the case as evident from Fig.\ref{1fsresult}. Here we plotted $\triangle^{\text{hbd}}$ as a function of $\tilde{v}$ at two different values of the dimensionless chemical potential $\tilde{\mu}_{b}$.  We find that the pairing gap increases when $\tilde{v}$ is reduced, reaches a peak, and then decreases to zero in the flat band limit when chemical potential $\tilde{\mu}_b = 0.50$ and $\tilde{\mu}_b = 0.75$. But when chemical potential is very low($\tilde{\mu}_b = 0.25$), the peak is reached only when $\tilde{v}\approx 0$. 
 
 This rather surprising result has to do with the renormalization factor in the interaction constant. It essentially gives the probability amplitude of the given electronic state to be in the thin film side of the interface. Its definition is given in Eqn.\ref{Zfactor}. Here $Z_3$ is defined as the renormalization factor of the electrons on the Fermi surface. Since the Dirac point is in resonance with the thin film transverse band, $Z_3$ is exactly $1/2$ at $k = 0$. But if the Fermi momentum is much greater than zero, then the renormalization factor changes from $1/2$. This is equivalent to detuning away from resonance. 
 If the hybrid band is adiabatically connected to the thin film band at large $k$, then $Z_3 \rightarrow 1$ at large Fermi momentum. On the other hand, if the hybrid band is connected to the surface Dirac cone, then $Z_3 \rightarrow 1$ at large Fermi momentum. This change in the renormalization factor can substantially affect the magnitude of the SC gap. 

 So what we observe here essentially is an interplay between the density of states at the Fermi level and the renormalization factor of the electronic states on the thin film side. The density of states increases with decreasing $\tilde{v}$ in a monotonic fashion for any value of $\tilde{\mu}_b$. This is evident from the density of states plot in Fig.\ref{1fsresult}(b). The density of states increases in a power law fashion in both cases of chemical potential as the $\tilde{v}$ is lowered. 
 
 On the other hand, the renormalization factor $Z_3$ decreases as $\tilde{v}$ is lowered(see Fig.\ref{1fsresult}(c)). This can be explained in the following way: Here the Fermi level crosses the positive helicity branch of the bottom band(band index - $(b,+)$). Consider the large $\tilde{v}$ limit, which is defined as the limit when $Z_3(\tilde{v}) > 1/2$. In this limit, the electrons in this band are adiabatically connected to the thin film band at a large $k$ limit, where they are out-of-resonance. So if the Fermi level crosses this band at large $k$, then $Z_3 \approx 1$. Also, notice that the range of momentum states around the Dirac point which experience strong hybridization decreases as $\tilde{v}$ is increased. As a result of these two factors, one can see why $Z_3$ increases when $\tilde{v}$ is increased. On the other hand, in the limit of $\tilde{v}$ when $Z_3(\tilde{v}) < 1/2$, the hybrid band under consideration(band index - $(b,+)$) is adiabatically connected to the non-interacting surface Dirac cone. This is the reason why $Z_3 \rightarrow 0$ as $\tilde{v} \rightarrow 0$. At $Z_3(\tilde{v}) = 1/2$, the electrons in the Fermi surface are in quantum-well resonance.

 The variation in $Z_3$ will be more substantial for cases with higher chemical potential than those with lower ones. Due to the higher chemical potential, the Fermi electrons are detuned away from resonance and hence the $Z_3$ factor will be different from $1/2$. This is the reason why we see a peak in the pairing gap for $\tilde{\mu}_b = 0.50, 0.75$(fig.\ref{1fsresult})(a). On the other hand,  $Z_3 \sim 1/2$ for $\tilde{\mu}_b = 0.25$ at all values of $\tilde{v}$, implying that the electrons lying in the Fermi surface are in quantum-well resonance throughout. As a result, the monotonic behavior of the density of states $\mathcal{N}^{\text{hbd}}$ is also reflected in the evolution of the pairing gap.  
 
\begin{figure*}
 \includegraphics[width=12.25cm, height=9cm]{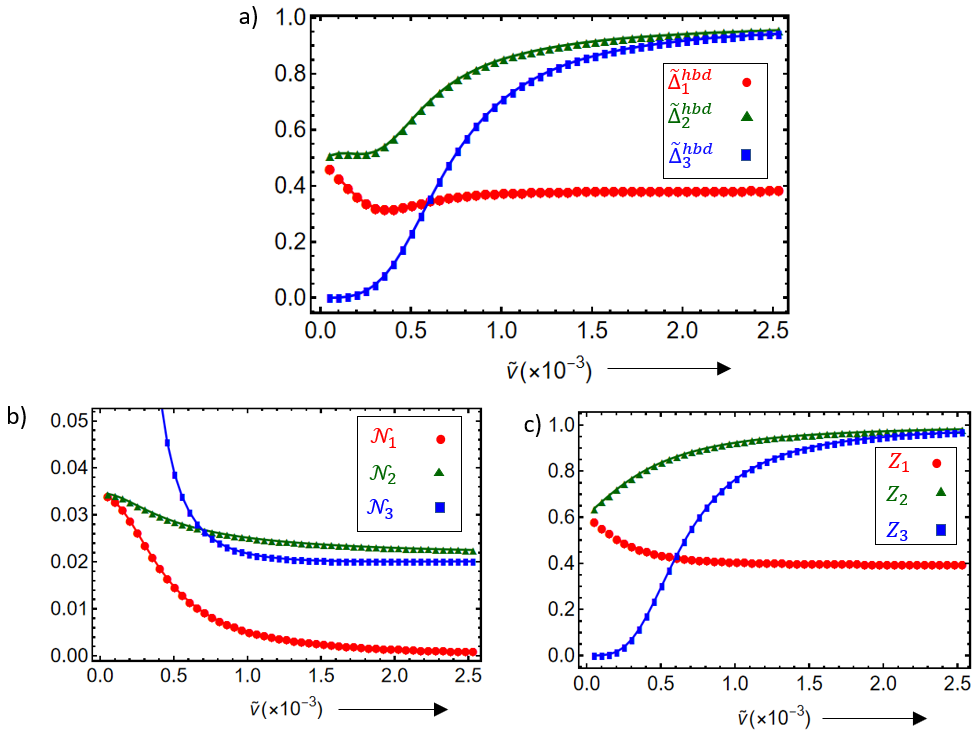} \\
 \caption{Three Fermi surface model(see fig.\ref{3fsspectra}): Similar setup as in Fig.\ref{1fsresult} but here the Fermi level is set to three Fermi surfaces. The effective chemical potential is fixed at $\tilde{\mu}_b = 2.25$. In (b), we plotted the density of states at the Fermi level of the three Fermi surfaces as a function of $\tilde{v}$. The exact definition of $N_i$(i=1,2,3) is given in Eqn.\ref{dos3fs}. In (c), we plotted the $Z$-factors of the Fermi electrons on the three Fermi surfaces.}
 \label{3fsn=1}
\end{figure*}

 \paragraph{Three Fermi surfaces}Here we study the evolution of the p-wave pairing gaps on the three FSs as a function of $\tilde{v}$. Just like in the previous case of a single FS, the tunneling strength and the thin film's material parameters are kept fixed.  Note that since the tunneling results in a 2-level splitting of the top and bottom bands at $k=0$ by a factor of $2t_d$, the effective chemical potential $\mu_b$(defined in Eqn.\ref{mudef}) should be of the order of $2t_d$ or greater than that to realize a three Fermi surface model. In other words, the dimensionless parameter $\tilde{\mu}_{b} \ge 2$. See the energy spectrum in fig.\ref{3fsspectra} for details.

In our calculations, we fix  
 the effective chemical potential at $\tilde{\mu}_b = 2.25$. We fix the tunneling strength at $t_d = 0.2 \text{eV}$. The results are shown in Fig.\ref{3fsn=1}(a). Here we plotted the p-wave superconducting gaps on the three FSs as a function of the SOC strength of the TI surface, represented in a dimensionless form as $\tilde{v}$ (defined in Eqn.\ref{soc}). As before, we numerically solved the self-consistent superconducting gap equations defined in Eqn.\ref{gapeq3fsn=1} to calculate the pairing amplitudes on the three Fermi surfaces. The pairing gaps have been represented in a dimensionless form by dividing it with the Debye frequency of the thin film metal. In Fig.\ref{3fsn=1}(b), we plotted the density of states at the Fermi level for each Fermi surface as a function of $\tilde{v}$. The definitions are given by,
\begin{eqnarray}
    \mathcal{N}^{\text{hbd}}_{1} &=& \int \frac{d^{2}\textbf{k}}{(2\pi)^2} \delta(\epsilon^{\text{hbd}}_{\textbf{k},t,+} - \mu)\nonumber\\  \mathcal{N}^{\text{hbd}}_{2} &=& \int \frac{d^{2}\textbf{k}}{(2\pi)^2} \delta(\epsilon^{\text{hbd}}_{\textbf{k},t,-} - \mu)\nonumber \\ \mathcal{N}^{\text{hbd}}_{3} &=& \int \frac{d^{2}\textbf{k}}{(2\pi)^2} \delta(\epsilon^{\text{hbd}}_{\textbf{k},b,+} - \mu)\label{dos3fs}
\end{eqnarray}
where $\mathcal{N}^{\text{hbd}}_{i}$(i=1,2,3) implies the density of states at the Fermi surface indexed by $i$ with $i=1$ being the closest to the Dirac point. In Fig.\ref{3fsn=1}(c), we plotted the renormalization factor $Z_{i}$(i=1,2,3) of the three Fermi surfaces. We studied the variation of the renormalization factors of the Fermi electrons on each Fermi surface as a function of $\tilde{v}$. 

Similar to what we saw in the single Fermi surface case, the magnitude of the SC gaps on the three Fermi surfaces is determined by the interplay of the electron density of states at the Fermi level and the renormalization factors $Z_i$. One can notice here by observing the Figs.\ref{3fsn=1}(a) and \ref{3fsn=1}(c) that it is the $Z$-factors in three FSs that play the dominant role here. To realize a three-Fermi surface model, we require $\tilde{\mu}_b \ge 2$. Thus the Fermi momentum of the 2nd and 3rd FSs are already much greater than zero. Thus the tunneling effect on these Fermi electrons becomes lesser and lesser significant as the spin-orbit coupling strength is tuned up, no matter what the absolute value of the tunneling strength is. In addition, we also notice that the two Fermi surfaces get closer with increasing $\tilde{v}$.  This is also reflected in the magnitude of the pairing gap. We find here that $|\tilde{\triangle}^{\text{hbd}}_{2}-  \tilde{\triangle}^{\text{hbd}}_{3}| \rightarrow 0$ as $\tilde{v}\rightarrow 1$. One can notice here that the triplet component of the pairing amplitude, which is proportional to the difference in the pairing amplitude on the positive and negative helicity branches for a given $\textbf{k}$, vanishes as a result.
Thus as $\tilde{v} \rightarrow 1$, the tunneling effect on the two Fermi surfaces is negligible, effectively leading to a trivial singlet pairing order on the two Fermi surfaces which essentially overlaps.
On the other hand, the electrons on the 1st Fermi surface have their Z-factor nearly equal to $1/2$, implying the electronic states are near resonance even if we increase $\tilde{v}$. This is because the Fermi momentum is very close to zero. But notice here that the density of states $\mathcal{N}_{1}$ is nearly zero as $\tilde{v}$ is increased. This implies that the superconducting gap is dominated by the scattering of Cooper pairs from the other two Fermi surfaces, rather than the intra-band scattering.    

When $\tilde{v}$ is decreased, we are effectively moving toward the flat band limit of the TI surface. The density of states at each hybrid Fermi surface shows a monotonic increase as expected. However, this is not reflected in the SC gap magnitude. We find here that the pairing amplitude on the third Fermi surface vanishes in the limit $\tilde{v} \rightarrow 0$. On the other hand, the pairing amplitudes on the first and the second Fermi surfaces converge. That is, we observe that $|\tilde{\triangle}^{\text{hbd}}_{1}-  \tilde{\triangle}^{\text{hbd}}_{2}| \rightarrow 0$ as $\tilde{v}\rightarrow 0$. This implies that the two Fermi surfaces overlap to form the trivial thin QW band and the superconductivity on them will turn out to be of the trivial s-wave order. Since the superconductivity on the third Fermi surface vanishes as $\tilde{v} \rightarrow 0$, the topological superconductivity is absent in the flat band limit.

So in conclusion, we explored the evolution of the pairing gaps as a function of the SOC strength on the three Fermi surfaces at a fixed chemical potential and tunneling strength. We found that in the limit of large $\tilde{v}$($\tilde{v}\rightarrow 1$), the second and the third Fermi surfaces overlap and the pairing on them is of spin-singlet order. The SC pairing on the innermost Fermi surface still maintains the p-wave character. Thus the topological character is still maintained. In the limit when $\tilde{v} \rightarrow 0$, we found that the electrons in the 3rd Fermi surface lie entirely on the TI surface side. Hence they are effectively non-interacting. The first and the second Fermi surfaces overlap and we effectively have singlet pairing superconductivity on them. Hence in the flat band limit, the hybrid is no longer topological.

\section{\label{sec:level7}The large $N$ limit}
Here we consider the situation when the thin film band which is in quantum-well resonance with the surface Dirac point has its band index $N$ very much greater than one. Physically, this limit can be realized by increasing the thickness of the thin film. This is because, the energy difference between the successive quantum well bands, $ |\epsilon_{\textbf{k},n} - \epsilon_{\textbf{k},n-1}| \propto 1/d^2$. In this situation, given that the Fermi level is adjusted close to the Dirac point, there will be $N - 1$ off-resonance degenerate thin film bands crossing the Fermi level. Hence after hybridization, we shall have $2N - 2$ off-resonance Fermi surfaces plus one or three hybrid Fermi surfaces.  When $N \gg 1$, we anticipate that the dominant contribution to the superconducting gap on the hybrid bands is coming from the scattering of the singlet pair of electrons from the trivial thin film Fermi surfaces. The pairing between the helical fermions of the hybrid bands will only have a negligible effect on the pairing gap on off-resonance thin film bands in this limit. Effectively, one can describe this limit as equivalent to an external s-wave pairing field acting on the hybrid bands. So this is similar to the well-known superconducting proximity effect but in the momentum space.

In the first part, we shall derive an analytical expression for the pairing gap on the hybrid Fermi surface(s) by employing the large $N$ approximation.
Using this, we essentially study how far the interaction between the hybrid fermions can enhance the superconducting gap on the hybrid Fermi surface. 

In the last part of this section, we show that the momentum space proximity effect smoothly transforms into the real space proximity effect in the perturbative limit of tunneling. The surface interaction only gives a higher-order correction to the proximity-induced superconducting gap. 
\subsection{Momentum space proximity effect}

Consider the case when the Fermi level is adjusted such that it crosses just a single hybrid Fermi surface. So we have $2N - 2$ off-resonance Fermi surfaces and one hybrid Fermi surface. The exact gap equation in the limit when $\omega_D \ll \mu$ is given in Eqns.\ref{gapeqnormal1FS}, \ref{gapeqtsc1FS}. In the large $N$ limit, we could make substantial simplifications to
arrive at an analytical expression. Recall that in all our calculations, we considered the attractive interaction in the thin film to be mediated by confined phonons as explained in section \ref{sec:level3}A. But as $N\rightarrow \infty$ which is attained by increasing the film thickness, it is a good approximation to replace the confined phonons with the bulk phonons. This essentially makes the interaction potential $V^{n,n'}_{\textbf{k},\textbf{p}}$ isotropic. In the limit when the thickness $d\rightarrow \infty$, the interaction potential is defined in Eqn.\ref{V} attains the following isotropic form,  $$V^{n,n'}_{\textbf{k},\textbf{p}} \approx \frac{G^{2}_{\text{fp}}}{d}\left(1 + \frac{\delta_{n,n'}}{2}\right)\theta(\omega_D - \xi^{\text{tf}}_{\textbf{k}})\theta(\omega_D - \xi^{\text{tf}}_{\textbf{p}})$$ where $\delta_{n.n'}$ here is the Dirac-delta function. Since the interaction potential is isotropic, the superconducting gap will also turn out to be the same on all the thin film QW bands.  Now we shall plug this back into Eqn.
\ref{gapeqnormal1FS}. Also in the large $N$ limit,  scattering of Cooper pairs from the hybrid Fermi surface will have only a negligible effect on the s-wave thin film superconducting gap. This means the second term in the LHS of Eqn.\ref{gapeqnormal1FS} is neglected. With all these approximations, we obtain the following simple analytical form for the thin film s-wave superconducting gap,
\begin{eqnarray}
    \triangle^{\text{tf}} &\approx& 2\omega_{D} \text{Exp}\left[-\frac{d}{G^2_{\text{fp}}\mathcal{N}^{\text{tf}}(N-1/2)} \right] \\ \text{where}\,\, \triangle^{\text{tf}}_{n} &=& \triangle^{\text{tf}}_{n'} = \triangle^{\text{tf}}, \forall n, n' \leq N  \nonumber
\end{eqnarray}
Here we used $\triangle^{\text{tf}}$ for the s-wave superconducting gap on the thin film bands. $\mathcal{N}^{\text{tf}} = \frac{m}{2\pi \hbar^2}$ is the density of states at the Fermi level of a thin film transverse band, given the electronic dispersion is quadratic. Now let us plug this back into the gap equation for the magnitude of the effective p-wave superconducting order parameter on the hybrid Fermi surface. After doing some algebra, we get,
\begin{eqnarray}
    \triangle^{\text{hbd}} &=& \frac{Z_3 \triangle^{\text{tf}}}{1 - \tilde{\lambda}^{\text{hbd}} \text{ln}\frac{2\omega_D}{\triangle^{\text{hbd}}}}\label{largeN}\\ \text{where}\,\,\tilde{\lambda}^{\text{hbd}} &=& \mathcal{J}^{3,3} \mathcal{N}^{\text{hbd}}\nonumber  
\end{eqnarray}
Here $\tilde{\lambda}^{\text{hbd}}$ is the dimensionless coupling strength of interaction between the helical hybrid fermions. $\mathcal{N}^{\text{hbd}}$ is the density of states at the hybrid Fermi surface. $\mathcal{J}^{3,3}$ defined in Eqn.\ref{J} is the renormalized interaction potential between the hybrid fermions. $Z_3$ in the numerator is the renormalization factor of the hybrid Fermi electrons(defined in Eqns.\ref{Zfactor}, \ref{Z}). This factor comes from the scattering matrix element $\mathcal{K}^{n,3}$ that determines the scattering of singlet pair of electrons from the off-resonance thin film Fermi surface to the hybrid Fermi surface. 
\begin{figure}[b]
\includegraphics[width=7.cm, height=5cm]{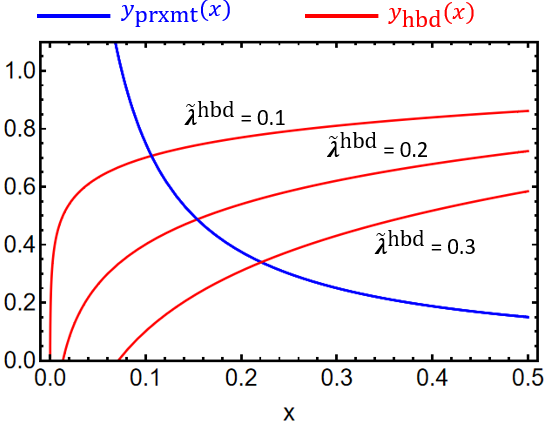}
\caption{\label{largeNlimit} An illustration of the enhancement effect on the superconducting gap on the hybrid Fermi surface due to the surface interaction. Here $x = \triangle^{\text{hbd}}/\omega_D$, the SC gap on the hybrid Fermi surface is taken as a variable. $y_{\text{prxmt}}(x)$(blue) takes into account the contribution to the pairing gap due to the scattering of Cooper pair from the off-resonance thin film Fermi surfaces. $y_{\text{hbd}}(x)$(red) is the contribution to the SC gap due to interaction between the hybrid fermions. The exact solution is at the point where the two curves cross each other.}
\end{figure}
\begin{figure}[b]
\includegraphics[width=7.cm, height=5cm]{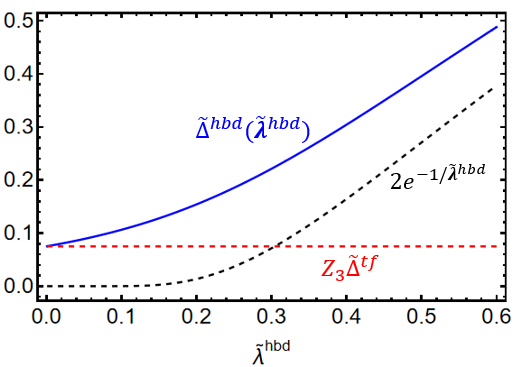}
\caption{\label{largeNlimitdelta}Here we solved the superconducting gap equation in the large-$N$ limit(Eqn.\ref{largeNlimit}) and plotted the SC order parameter $\triangle^{\text{hbd}}$(blue curve) as a function of the coupling strength between the hybrid fermions $\tilde{\lambda}^{\text{hbd}}$. The red dashed lines show the contribution to the SC gap due to the momentum space proximity effect. The black dashed line shows the solution solely due to the Cooper instability on the hybrid Fermi surface.} 
\end{figure}

Let us analyze the large $N$ result given in Eqn.\ref{largeN} more carefully. The numerator and the denominator come from different sources. The numerator is essentially the contribution to the superconducting gap due to the scattering of singlet-pair electrons from the off-resonance thin film Fermi surfaces. 
The denominator is due to the attractive interaction between the helical hybrid fermions. Hence it is this term that actually results in the Cooper instability on the hybrid Fermi surface. The numerator could open up a gap but does not lead to actual Cooper instability.

The numerator in the above expression is analogous to the proximity-induced superconductivity observed in several TI-SC heterostructures\cite{fukane(2008)}. In the proximity effect, the superconducting gap opens up on the Dirac cone due to the tunneling of Cooper pairs across the junction. The difference here is that the coefficient $Z$ here is nearly equal to $1/2$. In fact, as we shall demonstrate soon, the numerator turns out to be the proximity-induced superconducting gap in the perturbative limit of tunneling. The difference we notice in the resonance regime is that we observe an enhancement in the superconducting gap due to the attractive interaction between the helical hybrid fermions. The amount of enhancement is determined by the coupling strength $\tilde{\lambda}^{\text{hbd}}$. 

The fig.\ref{largeNlimit} below illustrates this enhancement effect on the p-wave pairing gap due to the interaction between the helical fermions. To do this, we defined the following functions, 
\begin{subequations}
\begin{eqnarray}
    y_{\text{hbd}}(x) &=& 1 - \tilde{\lambda}^{\text{hbd}} \text{ln}\frac{2}{x} \\ y_{\text{prxmt}}(x) &=& \frac{Z_3\tilde{\triangle}^{\text{tf}}}{x} 
\end{eqnarray}    
\end{subequations}
Here we replaced $\triangle^{\text{hbd}}/\omega_D$ in Eqn.\ref{largeN} by a variable $x$. So $\triangle^{\text{tf}}$ is also represented in a dimensionless form as $\tilde{\triangle}^{\text{tf}} = \triangle^{\text{tf}}/\omega_D$. $ y_{\text{hbd}}(x)$ is the contribution to the pairing gap due to the interaction between hybrid fermions. $y_{\text{prxmt}}(x)$ is the contribution due to the momentum-space proximity effect. The actual value for $x$ is found by solving the equation $y_{\text{hbd}}(x) = y_{\text{prxmt}}(x)$. We shall call the actual solution by $x_0$. One can call the solution to the equation $y_{\text{prxmt}}(x) = 1$ as the proximity limit of the superconductivity. This would have been the actual solution if the coupling constant $\tilde{\lambda}^{\text{hbd}} = 0$. Then we plotted the function $y_{\text{hbd}}(x)$ at different values of the coupling constant $\tilde{\lambda}^{\text{hbd}}$ in fig.\ref{largeNlimit}. Here we find that as the coupling constant is increased, the crossing point moves farther away from the proximity limit. This shows strong evidence of enhancement in the superconducting order due to interaction between hybridized fermions

To further emphasize this enhancement effect due to interaction between the hybrid fermions, we solved the equation $y_{\text{hbd}}(x) = y_{\text{prxmt}}(x)$ and plotted the resulting superconducting order parameter magnitude $\tilde{\triangle}^{\text{hbd}}$ as a function of the hybrid coupling constant $\tilde{\lambda}^{\text{hbd}}$. The results are shown in fig.\ref{largeNlimitdelta}. Here the red dashed lines are the proximity limit of the superconductivity obtained by solving $y_{\text{prxmt}} = 1$, while the black dashed lines give the BCS limit of the hybrid FS given by $y_{\text{hbd}}(x) = 0$. The enhancement due to the surface interaction exists even in the weakly interacting limit. As $\tilde{\lambda}^{\text{hbd}}$ approaches unity, we find that the order parameter attains an exponential form.  

But there are practical limitations in enhancing $\tilde{\lambda}^{\text{hbd}}$ to strongly interacting limit. The interaction potential $\mathcal{J}^{3,3}$ is predetermined by the bulk coupling constant of the thin film. At resonance, it is of the form $\mathcal{J}^{3,3} = Z^2_3 V^{N,N} \approx V^{N,N}/4$, which means it is always less than $\mathcal{K}^{n,3}$ for any $n$. So the only tunable parameter is the density of states at the Fermi level given by $\mathcal{N}^{\text{hbd}}$. If the energy dispersion of the hybrid band is linear when it crosses the Fermi level, then $\mathcal{N}^{\text{hbd}} = \frac{\mu_b}{2\pi A^{2}_b}$(Refer to Eqn.\ref{lineardirac}). Ideally, one could tune down the SOC strength of the TI surface to enhance the surface interaction. But as we discussed in section \ref{sec:level6}B, reducing the SOC strength will detune the Fermi electrons away from resonance for a fixed chemical potential, driving the Fermi surface back to the perturbative limit of tunneling. In short, what we like to convey here is that there are practical limitations in increasing the coupling strength $\tilde{\lambda}^{\text{hbd}}$. So in the large $N$ limit, the dominant contribution to the superconducting gap on the hybrid Fermi surface comes from the momentum space proximity effect due to the off-resonance thin film bands. There is an enhancement due to the Cooper instability on the hybrid Fermi surface, but that is not very substantial compared to the proximity effect.

\subsection{Perturbative limit of tunneling: Connection to the Fu-Kane model}

Here we shall consider the perturbative limit of tunneling by detuning away from the quantum-well resonance of the TI-thin film hybrid. Our objective here is to show that the momentum space proximity effect discussed in the previous section transforms into the real-space superconducting proximity effect in the perturbative limit of tunneling. 

The perturbative regime is characterized by the limit $\tilde{\delta} \gg 0$. Here $\tilde{\delta}$ is the dimensionless detuning parameter at $k=0$ defined in Eqn.\ref{detun}. So for convenience, we shall define a new parameter to study the perturbative limit given by,
\begin{eqnarray}
    \tilde{t} = \frac{1}{\tilde{\delta}}
\end{eqnarray}
where we can call $\tilde{t}$ as the dimensionless tunneling strength. This quantity essentially gives the probability amplitude of an electronic state in the thin film side to tunnel to the TI surface and vice versa.

In the perturbative regime, the single-particle hybridization effects are negligible. This implies that we should treat the surface fermions and the thin film fermions separately. This is evident from the discussions we had in section IV regarding the Z-effect. There we saw that on tuning $\tilde{\delta} \rightarrow -\infty$, the top hybrid band transforms to the surface Dirac cone and the bottom hybrid band transforms to the thin film band. Correspondingly $Z^b$ approaches unity while $Z^t$ approaches zero. It happens the other way when $\tilde{\delta} \rightarrow +\infty$. 

Now we shall see how the expression for   $\triangle^{\text{hbd}}$ derived in the large $N$ limit at quantum-well resonance(see Eqn.\ref{largeN}) changes when detuned to the perturbative limit. We shall be studying the perturbative limit for the case when $\tilde{\delta} \geq 0$. But the qualitative conclusions do not change when $\tilde{\delta} \leq 0$ also. If the Fermi momentum of the surface Dirac cone is very small, then $Z_3$ is essentially equal to $Z^{b}$ defined in Eqn.\ref{Zdetun}. For clarity, let us rewrite the expression again here. When Fermi momentum of the surface Dirac cone $k_F \approx 0$,
\begin{eqnarray}
    Z_3 = Z^b(\tilde{\delta}) = \frac{1}{2}\left( 1 - \frac{\tilde{\delta}}{\sqrt{1 + \tilde{\delta}^{2}}}\right)
\end{eqnarray}
Now expanding $Z_3$ in powers of $\tilde{t}$, we arrive at,
\begin{eqnarray}
    Z_3 = \tilde{t}^2 + \mathcal{O}(\tilde{t}^4)
\end{eqnarray}
Thus $Z_3$ scales as $\tilde{t}^2$ in the perturbative limit of tunneling. Recall that the coupling strength $\tilde{\lambda}^{\text{hbd}}$ determines the interaction between the surface fermions. Since there is no hybridization in this limit, let us call $\tilde{\lambda}^{\text{hbd}}$ as $\tilde{\lambda}^{\text{surf}}$. This is to emphasize that the coupling constant determines the attractive interaction strength between the surface fermions. Since the interaction potential is proportional to the square of the $Z$ factor, we see that in the perturbative limit,   
\begin{eqnarray}
    \tilde{\lambda}^{\text{surf}} &=& \alpha \tilde{t}^4 
\end{eqnarray}
where $\alpha = V^{N,N}\mathcal{N}^{\text{surf}}$. Here $N$ is the index of the thin film band that is closest to the TI surface. $\mathcal{N}^{\text{surf}}$ is the density of states at the Fermi level of the surface Dirac cone. Plugging this back to the Eqn.\ref{largeN}, the expression for the superconducting gap at the surface Dirac cone when expanded in powers of $\tilde{t}$ has the form,
\begin{equation}
    \triangle^{\text{surf}} \approx  \tilde{t}^2 \triangle^{\text{tf}} \left[1 + \alpha \tilde{t}^4 \ln \frac{2\omega_D}{\tilde{t}^2\triangle^{\text{tf}}} + ....\right]
\end{equation}
It is straightforward to find out that the first term is exactly the gap opening on the Dirac cone due to the superconducting proximity effect. Since the first term is proportional to the square of the tunneling strength, it has the most dominating effect on the SC gap magnitude on the surface. The second term is the lowest order correction to the gap magnitude due to a possible Cooper instability on the TI surface. We can see here that it has only a negligible contribution to the SC gap opening in the weak tunneling limit. 

In conclusion, by tuning our effective theory to the perturbative limit of tunneling, we could make connections to Fu-Kane's proposal. The momentum-space proximity effect we discovered in the large $N$ limit at the resonance transforms smoothly to the real-space proximity effect in the perturbative limit of tunneling. We also found that even in the perturbative limit, there is still an effective attractive interaction between surface fermions mediated by the thin film phonons. But this effect is so weak that the dominant contribution to the superconducting gap at the TI surface comes from the proximity effect.

\section{\label{sec:level8}General $N$ dependence}
In the previous sections, we studied the superconducting phase of the TI-thin film hybrid in the two extreme limits of $N$, the $N = 1$ limit, and the large $N$ limit. 
Here we shall probe the superconducting order parameter on the hybrid Fermi surfaces as a function of $N$. Tuning $N$ is implemented by increasing the thin film thickness. For each $N$, the thickness is further fine-tuned so that the Dirac point of the TI surface is at quantum-well resonance with the $N$th band of the thin film. So essentially we are studying the thickness dependence of the superconducting order parameter when the hybrid is fine-tuned to quantum-well resonance.  
\begin{figure}[b]
\includegraphics[width=8.3cm, height=6.6cm]{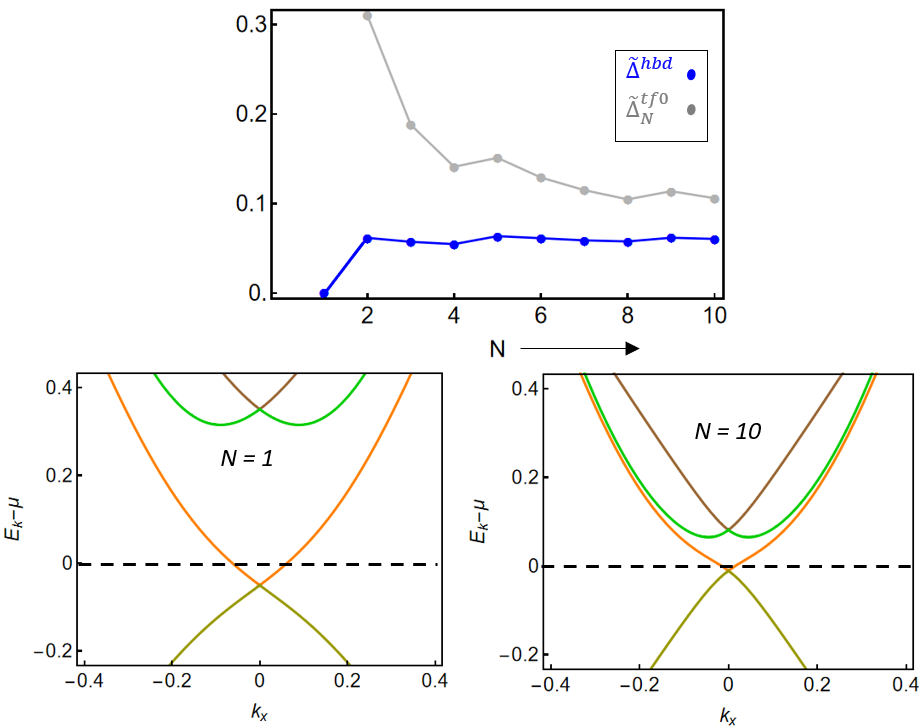}
\caption{\label{intermediateN1fs} Plot showing the evolution of the magnitude of SC order parameter on the single hybrid Fermi surface(Blue points) as a function of $N$. For each $N$, the thin film is at quantum-well resonance with the Dirac point of the TI surface. The effective chemical potential $\tilde{\mu}_b$(Defined in Eqn.\ref{effchempot}) is fixed at $\tilde{\mu}_b = 0.25$ for each $N$. $\triangle^{\text{tf0}}$ is the SC order parameter on the $N$th band of the thin film before the tunneling was switched on.  Also shown here are the energy spectra of the hybrid bands at the two limits, $N=1$ and $N=10$. The dashed lines represent the Fermi level. The spin-orbit coupling of the TI surface, $A_0 = 1.5 \text{eV}$\AA}
\end{figure}

Given that the hybrid is at quantum-well resonance for a given $N$, It is the following three quantities that would play a significant role as $N$ is tuned: thin film interaction potential matrix $V^{n,n,}$($n,n'$ are thin film band indices), the number of off-resonance thin film Fermi surfaces (equals $2N - 2$ for a given $N$) and the effective tunneling strength $t_d$. Recall from Eqn.\ref{V} that the thin film interaction potential scales as $1/d$ as a function of thickness. So even for a fixed bulk coupling constant $\tilde{\lambda}^{\text{bulk}}$, the interaction potential in the thin film decreases as a consequence of the electron confinement. But this is compensated by the increase in the number of bands that cross the Fermi level as thickness is tuned. This results in a jump in the superconducting order parameter each time a new band crosses the 
Fermi level. These two features have been studied extensively in the context of thin film superconductivity in previous works\cite{Sarma(2000)}. Recall from Eqn.\ref{Ht} that the electron confinement in the thin film leads to $1/\sqrt{d}$ scaling behavior of the tunneling strength. Thus, the effect of tunneling decreases with increasing thickness. Even though we would still see a splitting of the energy state at the Dirac point, the magnitude of the splitting substantially decreases at large $N$. Hence the evolution of the superconducting order parameters on the hybrid Fermi surface(s) as a function of $N$ will be a result of the interplay of these three factors.
We shall study the $N$ dependence for the single hybrid Fermi surface and three hybrid Fermi surfaces separately.

For numerical calculations, we used the material parameters of Pb for thin film. The spin-orbit coupling strength of the TI surface is fixed at $A_0 = 1.5 \text{eV}$\AA. 

\subsubsection{Single hybrid Fermi surface}
Fig.\ref{intermediateN1fs} shows the results when the Fermi level is tuned to one hybrid Fermi surface. Here the dimensionless effective chemical potential $\tilde{\mu}_b$(see Eqn.\ref{effchempot}) is fixed at $\tilde{\mu}_b = 0.25$. Note that fixing $\tilde{\mu}_b$ requires fine-tuning the Fermi level every time $N$ is increased. This is because the tunneling strength changes with thickness and $\tilde{\mu}_b = \mu_b/t_d$. So since we keep $\tilde{\mu}_b$ fixed, the absolute value of the chemical potential is not constant and changes with $N$. The p-wave superconducting gap on the hybrid Fermi surface for a given $N$ is found by solving the coupled self-consistent gap equation given in Eqns.\ref{gapeqtsc1FS},\ref{gapeqnormal1FS}  numerically. We calculated the SC order parameter value for $N$ values ranging from $1$ to $10$ by fine-tuning the thickness to quantum-well resonance for each $N$. Here $\triangle^{tf0}$(Grey) is the s-wave superconducting order parameter on the $N$th transverse band of the thin film  before the tunneling was turned on. This can be found easily using the same set of coupled equations by just setting tunneling strength to zero.

Here we find an enhancement in the gap magnitude as $N$ is increased from one. But from $N=3$ onwards, we find that the order parameter saturates to a constant value and it is a fraction of the thin film gap magnitude. This implies that the superconducting order on the hybrid Fermi surface approaches the large $N$ limit right from $N=2$ onwards. From our discussions in the previous section on the large $N$ limit, we can conclude that the superconducting order from $N=2$ onwards is dominated by the scattering of singlet pairs of electrons from the off-resonance thin film bands. So to conclude, at intermediate $N$ we find an enhancement in the pairing gap due to the off-resonance thin film Fermi surfaces which start appearing as $N$ is increased from one. At large $N$, the superconducting gap saturates to a constant value and is fixed by the thin film superconducting gap due to the momentum space proximity effect.

\subsubsection{Three hybrid Fermi surfaces}

\begin{figure}[b]
\includegraphics[width=8.3cm, height=6.6cm]{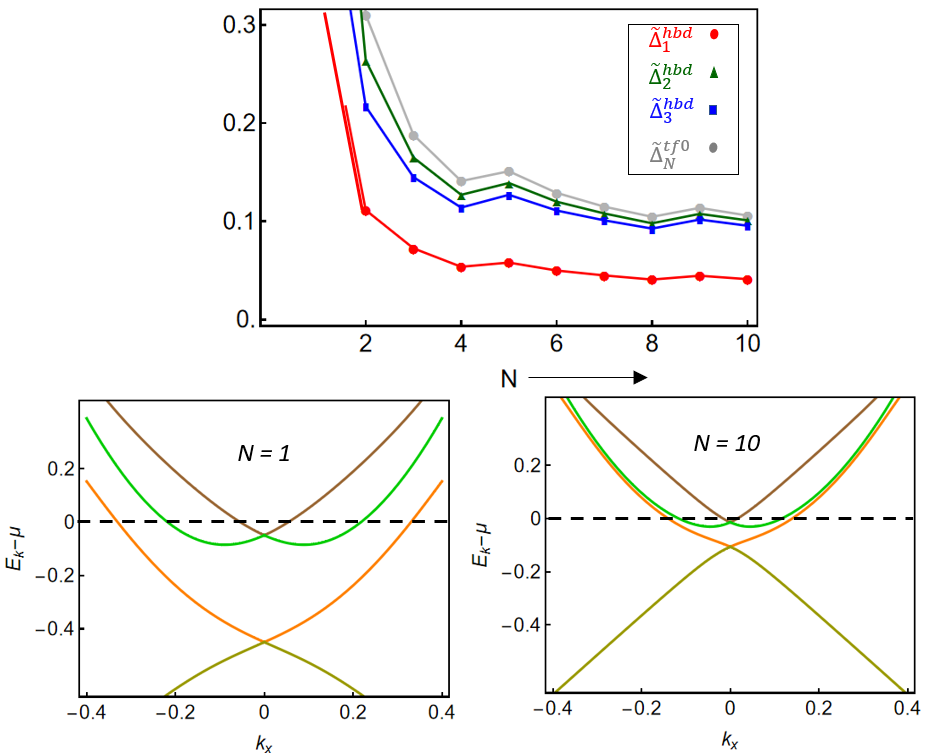}
\caption{\label{intermediateN3fs} Similar setting as in Fig.\ref{intermediateN1fs}. The difference here is that the effective chemical potential $\tilde{\mu}_b$ is set to $\tilde{\mu}_b = 2.25$ for all $N$. So we have three hybrid Fermi surfaces at quantum-well resonance. }
\end{figure}

Fig.\ref{intermediateN3fs} shows the results when the Fermi level is tuned to three hybrid Fermi surfaces limit. From our previous discussions on the three hybrid Fermi surfaces model in the $N=1$ limit, we understand that $\tilde{\mu}_b \geq 2$ to realize this model. Hence we set $\tilde{\mu}_b = 2.25$ for all $N$. We solve the coupled self-consistent equations given in Eqns.\ref{gapeqtsc},\ref{gapeqnormal} numerically for a given $N$ at quantum-well resonance. $\triangle^{\text{hbd}}_i$ gives the magnitude of the p-wave superconducting order parameter on the $i$th hybrid Fermi surface with $i=1$ being the closest one to the Dirac point. 

Unlike the single Fermi surface case, here the three superconducting order parameters decrease with increasing $N$ and then saturate to a constant value. Since we have three hybrid Fermi surfaces in the $N=1$ limit, the density of states is sufficiently high compared to the single Fermi surface case. So in this case, it is the $1/d$ scaling of the interaction potential that has a dominating effect on the superconducting order in the intermediate $N$ limit than the increase in the number of off-resonance bands.

As $N$ is increased, we observe that the superconducting gap on the second and the third Fermi surfaces start converging to the thin film gap value. This can be attributed to the
 $1/\sqrt{d}$ scaling of the tunneling strength. The tunneling gets weaker as $N$ is increased so that the electrons lying away from $k=0$ experience only a perturbative effect. This is evident from the energy spectrum of the hybrid bands in the $N=1$ and the $N=10$ limits shown in Fig.\ref{intermediateN3fs}.  As a result, the second and the third Fermi surfaces overlap and become degenerate. So the triplet component of the order parameter in the Zeeman basis cancels out and we are left with a trivial s-wave superconducting order on these two Fermi surfaces. In short, the two Fermi surfaces essentially became off-resonance.
But the pairing gap on the first Fermi surface is still of p-wave symmetry. Hence the hybrid is still in the topological phase.

So to conclude, the superconducting order parameter on the three hybrid Fermi surfaces decreases with increasing $N$ at intermediate values of $N$. This is a result of the $1/d$ scaling of the interaction potential. As $N$ is increased further, it is only the Fermi surface closest to the Dirac point that exhibits topological superconductivity. The other two Fermi surfaces which turn out to be at the off-resonance overlap and hence the superconducting order on them becomes trivial s-wave-like. 

\section{\label{sec:level9}Conclusion}
In this paper, we proposed a TI-thin film hybrid as a practical platform to realize a system with attractively interacting surface fermions. By depositing the thin film on top of the TI surface, we essentially allowed the surface electrons to be exported to the interacting thin film. We found that for a given thin film and the topological insulator, when the surface fermions resonate with the quantum-well states of the thin film, the interaction between surface fermions is maximally enhanced. 

Then we studied the superconductivity of these resonating hybrid states in the $N=1$ limit. In this limit, we effectively have a four-band model of interacting helical hybrid fermions. By fine-tuning the Fermi level in this limit, we showed that it is possible to construct an effective low-energy theory of a single flavor of 2-component Dirac fermions subject to attractive interaction, whose quantum critical point possesses emergent supersymmetry(SUSY). Then we studied the evolution of the superconducting gap  as a function of the interaction strength of the thin film and the effective speed of light of the surface fermions. We find an enhancement of the superconducting gap when the interaction strength is increased. On the other hand, the evolution of the superconducting gap as the TI surface is tuned to the flat band limit is rather non-monotonic. We showed that when the Fermi level is tuned to the single Fermi surface limit, as a result of the interplay between the density of states at the Fermi level and the renormalization factor in the interaction strength $Z_3$, the superconducting gap shows a peak at an intermediate value of $\tilde{v}$ and then dies off to zero in the flat band limit. But if the effective chemical potential $\tilde{\mu}_b \approx 0$, the peak is seen in the flat band limit.

We also showed that in the large-$N$ limit, the superconductivity of the resonating hybrid fermions is dominated by the scattering of the singlet pair of electrons from the off-resonance thin film bands. This effect is similar to the superconducting proximity effect but in the momentum space. However, interaction among the surface fermions can further enhance the superconducting gap.  In the strongly interacting limit of the surface, the enhancement effect can be very significant. 

We also studied the general $N$ dependence of the superconducting gap on the resonating helical hybrid bands. We found that when the Fermi level is tuned to three hybrid Fermi surfaces, the dominating effect is the $1/d$ scaling of the thin film interaction potential. The consequence of this scaling relation is that at resonance, the attractive interaction between the surface fermions is also at its maximum when $N = 1$. 

Apart from the theoretical interest in realizing a ground state of attractively interacting surface fermions, the proposed model also has practical applications in the context of Majorana-based quantum computation. Given that at resonance, the topological superconductivity is observed in the thin film side of the interface also, enhances the feasibility of experimental detection\cite{Trang2020}. Moreover, the amplitude of the superconducting order can be systematically adjusted by manipulating either the material's intrinsic properties or the geometric dimensions, as thoroughly discussed within the confines of this article. Such findings could pave the way for tangible advancements in quantum information technologies.

\appendix

\section{}
\label{deriv:HI}
First, let us project the Hamiltonian to the $d^{\dagger}_{\textbf{k},N,t(b)}\ket{0}$ states. 
This is made possible by the unitary transformation $d_{\textbf{k}} = U_{\textbf{k}}\Gamma_{\textbf{k}, N}$ given in Eqn.\ref{unitary}. Here the 2-component thin film spinor $c_{\textbf{k}, N}$ can be projected out of the 4-component $\Gamma$ using the relation $c_{\textbf{k},N} = \frac{1+ \sigma_{z}}{2} \Gamma_{\textbf{k}, N}$. Putting these two relations together, we get a relation connecting the $c$ basis with the $d$ basis. Then the singlet pair creation operator in the thin film basis $c^{\dagger}_{\textbf{k}, N}s_yc^{\dagger T}_{-\textbf{k}, N}$ transforms as:
\begin{widetext}
\begin{eqnarray}
    c^{\dagger}_{\textbf{k},N}(-is_y)c^{\dagger T}_{-\textbf{k},N} &=& d^{\dagger}_{\textbf{k}} U_{\textbf{k}}\frac{1+ \sigma_{z}}{2}(-is_y)\frac{1+ \sigma_{z}}{2}U^{T}_{-\textbf{k}}d^{\dagger T}_{-\textbf{k}}\nonumber\\ &=& \left(\begin{array}{cc}d^{\dagger}_{\textbf{k},t} & d^{\dagger}_{\textbf{k},b}  
    \end{array} \right) \left(\begin{array}{ccc}
        \cos^{2}\frac{\theta_{\textbf{k}}}{2} (-is_y)  & - \cos \frac{\theta_{\textbf{k}}}{2}\sin\frac{\theta_{\textbf{k}}}{2} (-is_y)\\  \\- \sin \frac{\theta_{\textbf{k}}}{2}\cos\frac{\theta_{\textbf{k}}}{2}  (-is_y)
         & \sin^{2}\frac{\theta_{\textbf{k}}}{2} (-is_y)
    \end{array}\right)\left(\begin{array}{c}d^{\dagger T}_{-\textbf{k},t} \\ d^{\dagger T}_{-\textbf{k},b}\end{array} \right)
\end{eqnarray}
\end{widetext}
where $d^{\dagger}_{\textbf{k},t(b)} = \left(\begin{array}{cc} d^{\dagger}_{\textbf{k},t(b), \uparrow} & d^{\dagger}_{\textbf{k}, t(b),\downarrow}
\end{array} \right)$ are the 2-component spinors in the spin-1/2 space representing the creation operators of the top(bottom) band. $\cos\frac{\theta_{\textbf{k}}}{2}$ and $\sin\frac{\theta_{\textbf{k}}}{2}$ are nothing but the projection of the 'top' and 'bottom hybrid states into the thin film state. That is,
\begin{eqnarray}
 \cos\frac{\theta_{\textbf{k}}}{2} = \bra{0}c_{\textbf{k},N}d^{\dagger}_{\textbf{k},t}\ket{0} \,\,\sin\frac{\theta_{\textbf{k}}}{2} = \bra{0}c_{\textbf{k},N}d^{\dagger}_{\textbf{k},b}\ket{0} 
\end{eqnarray}
 Remember that both these matrices have off-diagonal elements in the laboratory spin basis due to the induced spin-orbit coupling on these bands. The exact expression of $\cos\theta_{\textbf{k}}$ is given in Eqn.\ref{unitary}. 

The off-diagonal elements in the above matrix suggest the possibility of inter-band pairing. Since we are only interested in the weak pairing limit where only the pairing between the Fermi electrons is considered, the inter-band pairing does not occur in this limit. The weak-pairing approximation allows us to treat the pair creation operators for the top and bottom bands separately. Let us define $\hat{P}_t$ and $\hat{P}_b$ as the pair creation operators for the top and bottom bands respectively. We have,
\begin{subequations}
\begin{eqnarray}
    \hat{P}_{\textbf{k},t} &=& d^{\dagger}_{\textbf{k},t} \cos^{2}\frac{\theta_{\textbf{k}}}{2} (-is_y) d^{\dagger T}_{-\textbf{k},t} \\
    \hat{P}_{\textbf{k},b} &=& d^{\dagger}_{\textbf{k},b} \sin^{2}\frac{\theta_{\textbf{k}}}{2} (-is_y) d^{\dagger T}_{-\textbf{k},b}
\end{eqnarray}    
\end{subequations}
Due to the induced helical spin structure of the hybrid bands, the corresponding single-particle Hamiltonian is diagonal in the helicity basis. As we said before, in the weak-pairing limit, the study of interaction will be easier if we project the interaction Hamiltonian also into the helicity basis. To implement this, let us write down the unitary matrix in the spin-1/2 space that can rotate the coordinates from the laboratory spin basis to the helicity basis,
\begin{eqnarray}
    d^{\dagger}_{\textbf{k},t(b)} = a^{\dagger}_{\textbf{k},t(b)}\Pi^{\dagger}_{\textbf{k}}, && \Pi_{\textbf{k}} = \frac{1}{\sqrt{2}}\left(\begin{array}{cc}
        1 & 1 \\
        e^{i\phi_{\textbf{k}}} & -e^{i\phi_{\textbf{k}}}
    \end{array} \right)  
\end{eqnarray}

Now we shall plug this back into the set of pair creation operators defined above. Here we observe that the matrices $\cos^2 \frac{\theta_{\textbf{k}}}{2}$ and $\sin^2 \frac{\theta_{\textbf{k}}}{2}$ are diagonal in the helicity basis. This is because the only way an off-diagonal term can appear in these matrices is through the spin-orbit coupling term of the TI surface. With this information and after doing some algebra, we get,
\begin{subequations}
\begin{multline}
    \hat{P}_{\textbf{k},t} =\\ \left(\begin{array}{cc}
        a^{\dagger}_{\textbf{k},t,+} & a^{\dagger}_{\textbf{k},t,-} 
    \end{array} \right) \left(\begin{array}{cc}
        Z^{t}_{\textbf{k},+} & 0 \\
       0  & - Z^{t}_{\textbf{k},-}
    \end{array} \right)\left(\begin{array}{c}
        e^{-i\phi_{\textbf{k}}} a^{\dagger}_{-\textbf{k},t,+} \\ e^{-i\phi_{\textbf{k}}} a^{\dagger}_{-\textbf{k},t,-} 
    \end{array} \right)
    \end{multline}
    \begin{multline}
    \hat{P}_{\textbf{k},b} =\\ \left(\begin{array}{cc}
        a^{\dagger}_{\textbf{k},b,+} & a^{\dagger}_{\textbf{k},b,-} 
    \end{array} \right) \left(\begin{array}{cc}
        Z^{b}_{\textbf{k},+} & 0 \\
       0  & - Z^{b}_{\textbf{k},-}
    \end{array} \right)\left(\begin{array}{c}
        e^{-i\phi_{\textbf{k}}} a^{\dagger}_{-\textbf{k},b,+} \\ e^{-i\phi_{\textbf{k}}} a^{\dagger}_{-\textbf{k},b,-} 
    \end{array} \right)    
    \end{multline}
\begin{eqnarray}
    Z^{t}_{\textbf{k},\pm} &=& \frac{1}{2}\left( 1 + \frac{\delta_{\textbf{k},\pm}}{\sqrt{\delta^2_{\textbf{k},\pm} + \frac{t^2}{d}}}\right)\label{Z1}\\ Z^{b}_{\textbf{k},\pm} &=& \frac{1}{2}\left( 1 - \frac{\delta_{\textbf{k},\pm}}{\sqrt{\delta^2_{\textbf{k},\pm} + \frac{t^2}{d}}}\right)\label{Z2}\\ \delta_{\textbf{k},\pm} &=& \frac{1}{2}\left(\epsilon^{\text{tf}}_{\textbf{k},N} - \epsilon^{\text{surf}}_{\textbf{k},\pm} \right)
\end{eqnarray}
\end{subequations}
 .  

\nocite{*}
\bibliography{ref}

\end{document}